# The semiclassical resonance spectrum of Hydrogen in a constant magnetic field

Gregor Tanner*
Niels Bohr Institute, Blegdamsvej 17,
DK-2100 Copenhagen Ø, Denmark

Kai T. Hansen[†‡]
Fakultät für Physik, Universität Freiburg
Hermann-Herder-Strasse 3, D-79104 Freiburg, Germany
and
Jörg Main [§]
Institut für Theoretische Physik I, Ruhr-Universität Bochum,
D-44780 Bochum, Germany

December 12, 1995

## Abstract

We present the first purely semiclassical calculation of the resonance spectrum in the Diamagnetic Kepler problem (DKP), a hydrogen atom in a constant magnetic field with $L_z = 0$. The classical system is unbound and completely chaotic for a scaled energy $\epsilon \sim EB^{-2/3}$ larger than a critical value $\epsilon_c > 0$. The quantum mechanical resonances can in semiclassical approximation be expressed as the zeros of the semiclassical zeta function, a product over all the periodic orbits of the underlying classical dynamics. Intermittency originating from the asymptotically separable limit of the potential at large electron–nucleus distance causes divergences in the periodic orbit formula. Using a regularisation technique introduced in [G. Tanner and D. Wintgen, Phys. Rev. Lett. **75**, 2928 (1995)] together with a modified cycle expansion, we calculate semiclassical resonances, both position and width, which are in good agreement with quantum mechanical results obtained by the method of complex rotation. The method also provides good estimates for the bound state spectrum obtained here from the classical dynamics of a scattering system. A quasi Einstein–Brillouin–Keller (QEBK) quantisation is derived that allows for a description of the spectrum in terms of approximate quantum numbers and yields the correct asymptotic behaviour of the Rydberg–like series converging towards the different Landau thresholds.

---

*email: tanner@kaos.nbi.dk
[†]email: khansen@kaos.nbi.dk
[‡]Present address: NORDITA, Blegdamsvej 17, DK-2100 Copenhagen Ø, Denmark. Also at: Physics Department, University of Oslo, Box 1048, Blindern, N-0316 Oslo, Norway.
[§]email: main@tp1.ruhr-uni-bochum.de

The hydrogen atom in a uniform magnetic field has become one of the most important examples for studying the correspondence between quantum mechanics and classical chaos. The Hamiltonian is known to a high accuracy and furthermore, the system is experimentally accessible in the laboratory. The angular momentum in the direction of the magnetic field is conserved, which reduces the classical system to a problem with two degrees of freedom. The classical flow in phase space covers a wide range of Hamiltonian dynamics reaching from bound, nearly integrable behaviour to completely chaotic and unbound motion by varying one parameter, the scaled energy $\epsilon$.

In the 1980's, the system served as a catalyst for quantum chaos. Modulations in the absorption spectra of highly excited hydrogen atoms in a magnetic field, the so–called quasi Landau levels [1, 2], could be assigned to classical trajectories [3, 4] and could be understood in the framework of the semiclassical periodic orbit theory [5, 6, 7]. This strikingly simple answer to a long outstanding question led to a completely new viewpoint in analysing quantum spectra in general. We refer the reader to the review articles and article collections [8, 9, 10, 11, 12].

So far, the main interest in the literature laid on *understanding* quantum manifestations of the system such as the quantum spectra or the wave functions, in terms of the underlying classical mechanics. Effective numerical techniques to solve Schrödinger's equation directly (neglecting relativistic and finite mass effects) have been developed in the recent years, both for the bound state spectra [13, 14, 15, 16, 17] and resonances in the continuum [18, 19]. Periodic or closed orbits appear in the Fourier transformed energy – or photo ionisation – spectra obtained from experiments or from full quantum calculations. Classical orbits can in some cases be identified as "scars" in quantum eigenfunctions [20] and as prominent peaks in the corresponding Husimi distributions [21].

Less effort has been undertaken in the opposite direction, i.e. in quantising hydrogen in a magnetic field directly in terms of classical entities only. The density of quantum states [22], photo absorption spectra [23, 24] or the evolution of Rydberg wave packets [25] can formally be written as sum over periodic or closed orbits. A brute force application of the orbit sums using the shortest classical orbits only yields the coarse grained structure of the spectrum [26] or the time evolution of Rydberg wave packets on short time scales [25], respectively. To resolve the fine structure of the spectrum and individual highly excited quantum states, the information of the classical dynamics on long time scales is needed. The classical orbit sums, however, diverge due to the exponential growth of the number of orbits in chaotic systems and appropriate resummation techniques have to be developed to overcome this problem.

Starting point is here the spectral determinant $D(E) = \prod_n (E - E_n)$, which can in semiclassical approximation be written as product over all periodic orbits of the system, the so–called semiclassical zeta function [27]. An analytic continuation of the product formula may be given by expanding the product and regrouping terms with the help of a symbolic description of the flow in form of a cycle expansion [28, 29]. This method has been shown to work for strictly hyperbolic classical dynamics, where all periodic orbits are unstable and the Lyapunov exponents of the orbits, i.e. the logarithm of the largest eigenvalue of the Monodromy matrix divided by the period, are strictly bound away from zero.

The Diamagnetic Kepler problem (DKP) corresponding to conserved angular momentum $L_z = 0$, (with the magnetic field in $z$ direction), is known to be completely chaotic in the sense that all periodic orbits are unstable for values of the scaled energy $\epsilon > \epsilon_c = 0.328782\ldots$ [30]. The classical motion is unbound in this regime and the flow in phase space can be described in form of a complete ternary symbolic dynamics [31, 30]. The DKP is, however, not hyperbolic even for energies $\epsilon > \epsilon_c$. The potential becomes separable in the limit where the classical electron is far from the nucleus, causing regular, but unstable, dynamics in this phase space region. As a consequence, the Lyapunov exponents of periodic orbits extending far into the regular region of phase space tend to zero. This behaviour, known as *intermittency* in the chaos-literature, is typically for the neighbourhood of marginal stable orbits or stable islands and is thus generic in Hamiltonian systems.

The regular limit of the classical motion causes divergences in the semiclassical periodic orbit expressions. This kind of divergence is different from the well known convergence problem of semiclassical expressions due to the exponential growth of the number of periodic orbits. Divergences introduced through intermittency can be regularised by first summing over periodic orbit contributions in the regular classical regime alone. An analytic continuation of each of the various sums can be given explicitly [32]. This procedure is equivalent to a cycle expansion of the zeta function in terms of an infinite alphabet [29]. We can finally present the first purely semiclassical quantisation of the resonance spectrum of DKP at fixed $\epsilon > \epsilon_c$. The regular limit of the dynamics can be identify to cause the Rydberg series structure at the various Landau thresholds. The high excited Rydberg resonances can be well described by a modified Einstein–Brillouin–Keller quantisation.

The article is written self-consistently in such a way that it contains all the information needed to understand the classical, the quantum and the semiclassical aspects of the problem. The classical system together with the symbolic dynamics is introduced in section 1. Asymptotic expressions for the actions and stability exponents in the separable limit of the potential are derived here. In section 2, we explain the method of complex rotation, which is an efficient tool to treat the full quantum problem in cases of resonances in the continuum. In the last section, we present our new semiclassical quantisation, and results are compared with full quantum calculations.

# 1 Classical dynamics

The non relativistic classical Hamiltonian for the hydrogen atom in a uniform magnetic field with field strength $B$ along the $z$–axis is given as

$$H = \frac{\mathbf{p}^2}{2m_e} - \frac{e^2}{r} + \frac{1}{2}m_e\omega^2(x^2+y^2) + \omega L_z. \tag{1}$$

The $z$ – component of the angular momentum $L_z$ is conserved and we will restrict ourself to the problem $L_z = 0$ in the following. We work in the infinite nucleus mass approximation and $m_e$ denotes the mass of the electron. The frequency $\omega = eB/2m_e c$ is half the cyclotron frequency. The Hamiltonian (1) in atomic units and for $L_z = 0$ has the form

$$H = E = \frac{p_z^2}{2} + \frac{p_\rho^2}{2} - \frac{1}{\sqrt{\rho^2+z^2}} + \frac{1}{8}\gamma^2\rho^2 \tag{2}$$

with the magnetic field strength $\gamma = B/B_0$ written in units of $B_0 = m_e^2 e^3 c/\hbar = 2.35 \cdot 10^5$ T. The radial distance from the $z$–axis is given by the coordinate $\rho$. The system is bounded for $E < 0$, while for $E > 0$ almost all trajectories escape to $z = \pm\infty$ with nonzero kinetic energy.

Introducing the scaling transformation

$$\mathbf{r} = \gamma^{-2/3}\tilde{\mathbf{r}}, \ \mathbf{p} = \gamma^{1/3}\tilde{\mathbf{p}}, \tag{3}$$

yields the new Hamiltonian

$$\tilde{H} = \epsilon = \gamma^{-2/3}H = \frac{\tilde{p}_\rho^2}{2} + \frac{\tilde{p}_z^2}{2} - \frac{1}{\sqrt{\tilde{\rho}^2+\tilde{z}^2}} + \frac{1}{8}\tilde{\rho}^2, \tag{4}$$

which is independent of $\gamma$. The classical dynamics is now controlled by one parameter only, the scaled energy $\epsilon = E\gamma^{-2/3}$. The classical action along a trajectory scales with the magnetic field like

$$S = \int \mathbf{p}\, d\mathbf{q} = \gamma^{-1/3}\tilde{S} \tag{5}$$

with $\epsilon$ fixed.

The Hamiltonian (4) is singular at $\tilde{\mathbf{r}} = 0$ yielding singular equations of motions at this point. For solving the equations of motion numerically, it is more convenient to switch to semi-parabolic coordinates and momenta [8]

$$\nu^2 = \tilde{r} - \tilde{z}, \ \mu^2 = \tilde{r} + \tilde{z}, \ p_\nu = d\nu/d\tau, \ p_\mu = d\mu/d\tau, \tag{6}$$

with a new rescaled time $\tau$

$$d\tau = dt/2\tilde{r} = (\nu^2+\mu^2)^{-1}dt. \tag{7}$$

The Hamiltonian (4) in semi-parabolic coordinates is

$$h = \frac{1}{2}p_\nu^2 + \frac{1}{2}p_\mu^2 - \epsilon(\nu^2+\mu^2) + \frac{1}{8}\nu^2\mu^2(\nu^2+\mu^2) \equiv 2, \tag{8}$$

and the scaled energy $\epsilon$ enters as a parameter here.

The structure of the dynamics of the Hamiltonian (8) depends on the value of the scaled energy $\epsilon$ alone. The $E$–$\gamma$ parameter plane can be partitioned into 5 distinct regions [8, 30], see also Fig. 1:

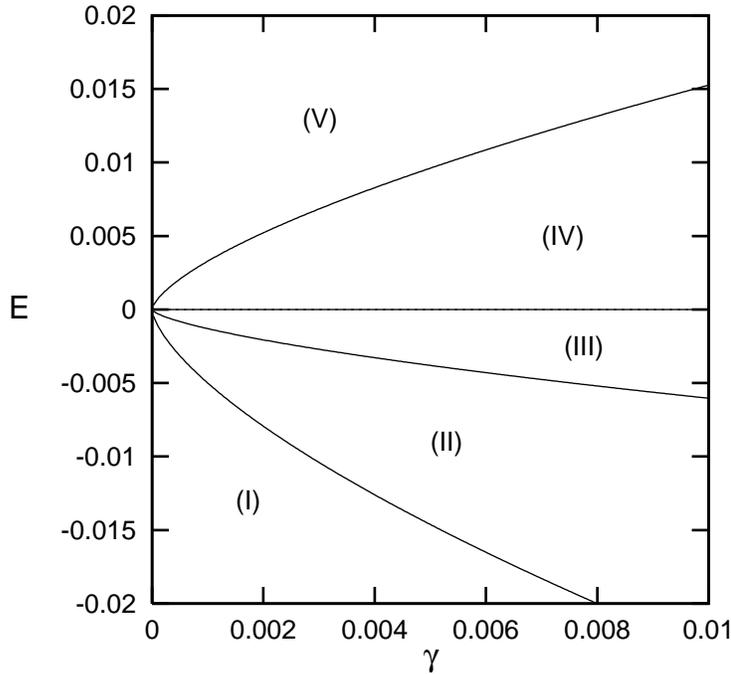

Figure 1: The 5 main dynamical regions in the $E$–$\gamma$ parameter space reaching from bound and almost integrable dynamics (I) to unbound completely chaotic dynamics (V). See also the explanation in the text.

(I) $\epsilon < -0.5$: bounded almost integrable motion;

(II) $-0.5 < \epsilon < -0.13$: bounded motion with mixed chaotic and regular motion;

(III) $-0.13 < \epsilon < 0.0$: the last large stable island disappear; the dynamics is mostly chaotic;

(IV) $0.0 < \epsilon < 0.328782\ldots$: unbounded mostly chaotic motion; the symbolic dynamic is not complete;

(V) $0.328782\ldots = \epsilon_c < \epsilon$: unbounded chaotic motion with a complete symbolic description.

In the following, we will always work in the region $\epsilon > \epsilon_c = 0.32878\ldots$, where a simple and complete symbolic description can be assigned to flow in phase space [30]. Furthermore, all periodic orbits are unstable in this parameter regime. For $\epsilon$ values below the critical energy $\epsilon_c$, the symbolic dynamics become incomplete and stable islands appear. Stable classical motion implies additional complications in the semiclassical description in section 3, which will not be included in the work presented here.

## 1.1 Symbolic dynamics and periodic orbits

We can motivate the structures of the classical motion and the underlying symbolic description by first studying the dynamics of a symmetric four disk scattering billiard [33]. The similarity between the DKP and the four disk system becomes obvious, when looking at the shape of the potential in the regularized $\mu, \nu$ – coordinates, see Fig. 2 (b). In the four disk billiard, a trajectory may after a bounce with one disk either hit one of the other three disk or escape from the system. A trajectory bouncing two times after leaving the first disk may bounce in 9 different ways. We find that there are $3^n$ possible sequences of disks for $n$ bounces. This set of non-escaping starting points for $n \to \infty$ is called a Cantor set. We label each disk with a number $s_t \in \{1, 2, 3, 4\}$ and a string $s_1 \cdots s_n$ assigned to a trajectory represents one of the $3^n$ possible sequences. A particle cannot bounce twice off the same disk, which is the only restriction in the 4–letter symbolic dynamics as long as the distance between the disks is sufficiently large compared to the disk radius. The orbits remaining in the system both forward and backward infinitely long in time form a two dimensional Cantor set and are uniquely described by bi–infinite symbol strings $\cdots s_{-2} s_{-1} s_0 s_1 s_2 \cdots$.

Since the Cantor set is ternary we can reduce the number of symbols to three with no restrictions on the sequence of symbols. The reduction can be done in several possible ways [31, 33]. In Appendix A, we construct a ternary alphabet closely related to the $C_{4v}$ – symmetry of our system, which will be used throughout the paper. A trajectory is now characterized by a symbol string $\cdots g_{-1} g_0 g_1 \cdots$ with $g_i \in \{0, 1, 2\}$ defined in Appendix A.

In Appendix B, we introduce a so called well ordered symbolic dynamics $\cdots w_{-1} w_0 w_1 \cdots$ with $w_i \in \{0, 1, 2\}$, which is in particular useful for finding periodic orbits numerically [34].

The structure of the Cantor set is similar for the DKP and for the four-disk scattering system. We can associate a 4–letter symbol string to every trajectory also in the DKP, and the symbols can be understood as a smooth "bounce" with one of the steep hills in the potential (8). At this bounces, a trajectory has a caustic, and we can determine the symbolic dynamics either by determining the number of caustics [31] or by making a suitable partition curve in a Poincaré plane [35]. A detailed discussion of this method will be given elsewhere [36].

The symbolic dynamics is complete, i.e. each possible symbol string $\cdots g_{-1} g_0 g_1 \cdots$ corresponds to a non-escaping orbit in the symmetric four disk system, if and only if the gap between two disks are larger than 0.205 times the disk radius[37]. If the disks come closer to each other certain paths between the disks are forbidden. These paths lie now in the "shadow" of a disk in between. The corresponding symbol strings cannot be related to a physical orbit and the symbolic dynamics is said to be pruned [38, 39, 37].

We find an analogous situation in the DKP. Here, each possible symbol string corresponds to an orbit if the scaled energy is larger than $\epsilon_c = 0.32878 \ldots$ [30]. For energies below the critical value, the hills of the potential will shadow some of the trajectories which exist for larger $\epsilon$ and the symbolic dynamics becomes pruned. The value of $\epsilon$ thus plays in the DKP the same role as the distance between the disks in the 4-disk system. Note that pruning in the DKP is introduced through bifurcations and stable islands appear and disappear with varying $\epsilon < \epsilon_c$. In the disk system, however, periodic orbits are always unstable and disappear immediately at a bifurcation point.

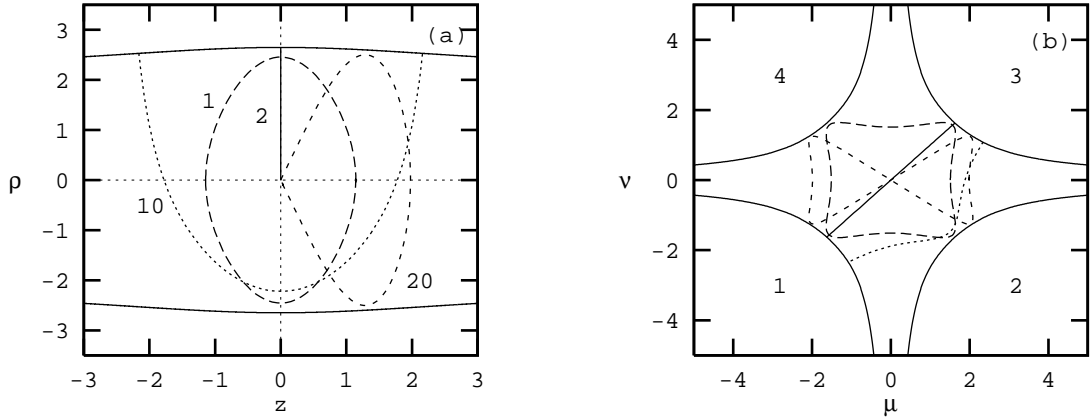

Figure 2: Periodic orbits in the (a) $(z, \rho)$ plane and (b) $(\nu, \mu)$ plane. The full line denotes the boundary of the potential for $\epsilon = 0.5$.

The periodic orbits are a subset of all non-escaping orbits existing in the system, (i.e. of the repellor). A periodic orbit is described by an infinite repetition of a finite symbol string $G = g_1 g_2 \cdots g_n$. In the following, we focus mainly on periodic orbits, which we characterize by the shortest non repeating symbol string directly.

Four short periodic orbits of the DKP are drawn in Fig. 2, both in the coordinate space $(z, \rho)$, and in the semi parabolic coordinates $(\nu, \mu)$. Note that the two orbits labelled $G=2$ and $G=20$ in Fig. 2 (a) collide with the nucleus. In Fig. 2 (b) we have indicated the labelling of the four hills used to construct the 4–letter symbolic dynamics. A translation from the 4 – letter to the 3 – letter alphabet can now be read off directly with the help of Appendix A. The orbits labelled 1, 2, 10, and 20 in Fig. 2 (a) are given the symbolic descriptions $S = 1234, 13, 1232$, and $1324$, respectively, in the 4 – letter code, Fig. 2 (b).

## 1.2 Asymptotic in the classical Landau channel

In this section, we will study the dynamics of the Hamiltonian (2) for positive energy $E$ in the limit $|z| \to \infty$. The coupling between the $z$ and $\rho$ coordinates vanishes in that limit and the Hamiltonian (2) has the asymptotic form

$$H = \frac{1}{2}p_\rho^2 + \frac{1}{2}p_z^2 + \frac{\omega^2}{2}\rho^2 - \frac{1}{|z|} + \mathcal{O}(\rho^2/z^3) = E > 0, \qquad (9)$$

with $\omega = \gamma/2$. The motion separates for large $|z|$ values in a pure Coulomb part along the $z$ – axis and an harmonic oscillator perpendicular to the magnetic field. The energies

$$E_h = \frac{1}{2}p_\rho^2 + \frac{\omega^2}{2}\rho^2 \quad \text{and} \quad E_c = \frac{1}{2}p_z^2 + \frac{1}{|z|} \qquad (10)$$

become adiabatic constants of motion for $|z|/\rho \to \infty$. The regular region far from the nucleus is always coupled to the chaotic motion near the core due to the attractive Coulomb force in $z$ – direction. The classical motion of the electron in the DKP thus alternates between strong chaotic motion for $\rho \approx |z|$ and regular time intervals out in the channel $|z| \gg \rho$. Almost all trajectories escape finally to infinity with nonzero momentum in $z$ – direction. Orbits which go further and further out in the regular channel but return to the origin, approach a marginally stable periodic orbit $|z| \equiv \infty$, $p_z \equiv 0$. In the symbolic description explained in the appendix A, this marginally stable orbit is labelled 0. Escaping trajectories which continue travelling along the $z$ – axis can be separated from returning once by a plane, that is part of the stable manifold of the marginal stable orbit. This plane is in the asymptotic limit given by the condition $E_c = p_z^2/2 - 1/|z| = 0$. Returning trajectories can be associated with bound motion in the $z$ – direction and $E_c < 0$.

Periodic orbits that go far out in the regular region $|z| \gg \rho$ pick up regular contributions to the actions and stability exponents, which can be given in analytic form to leading order. In the ternary symbolic description the channel orbits are characterized by a long string of consecutive symbols '0' in the symbol code. Of particular interest are periodic orbits with a symbol code of the form $G0^n$, i.e. periodic orbits with a common head string $G$ and a tail of $n$ symbols '0'. We will call these orbits a periodic orbit family. The orbits in a family have approximately the same behavior before entering and after leaving the regular region. They differ only in the '0' – tail string, i.e. in the number $n$ of half – oscillations perpendicular to the $z$ – axis in the regular region. In Fig. 3, some members of the periodic orbit families with head string $G = 1$ are shown explicitly.

The leading $n$ – dependent behavior in the actions and stability exponents is universal, i.e. it does not depend on the past or future of the periodic orbits before entering and after leaving the regular phase. Non universal contribution approaches a constant for orbits with the same head string $G$, but increasing number of oscillations in the regular regime.

To obtain the leading contributions to the actions, we start with the Hamiltonian (9) written in action–angle variables $(J_h, \varphi_h, J_c, \varphi_c)$,

$$H_{sep} = 2\omega J_h - \frac{1}{2J_c^2} \tag{11}$$

and omitting the coupling terms. The action variables are given as

$$J_h = \frac{1}{2\pi} \oint p_\rho d\rho = \frac{1}{2\omega} E_h \tag{12}$$

$$J_c = \frac{1}{2\pi} \oint p_z dz = \frac{1}{\sqrt{-2E_c}}. \tag{13}$$

The integration is taken over a single revolution of a trajectory in the potential of the one dimensional Hamiltonians in (10). (Note, that $\rho \geq 0$ in (12)). The angle variables propagate linearly in time with frequencies

$$\omega_h = \frac{\partial H_{sep}}{\partial J_h} = 2\omega; \quad \omega_c = \frac{\partial H_{sep}}{\partial J_c} = J_c^{-3}. \tag{14}$$

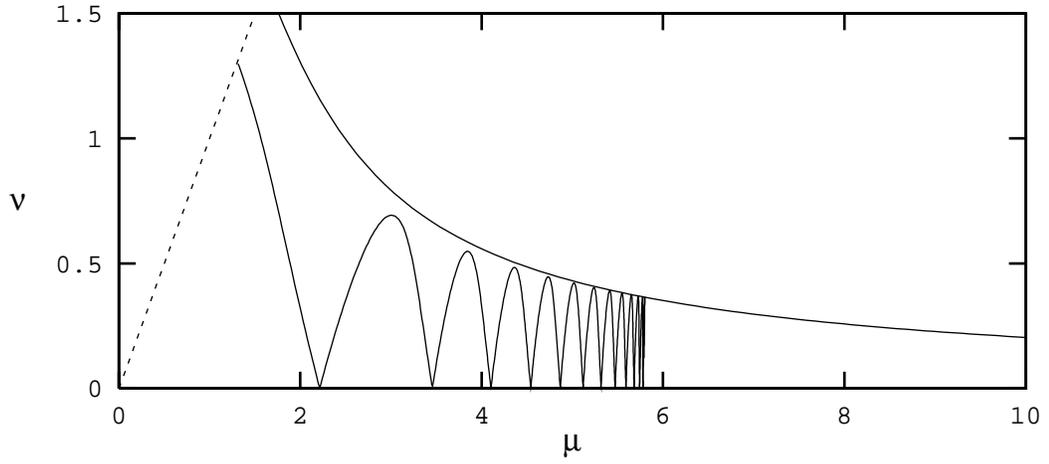

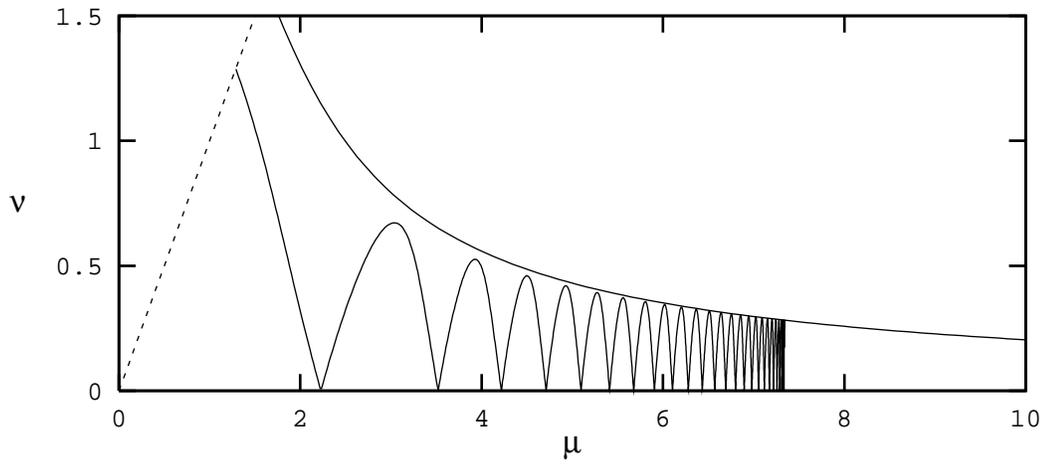

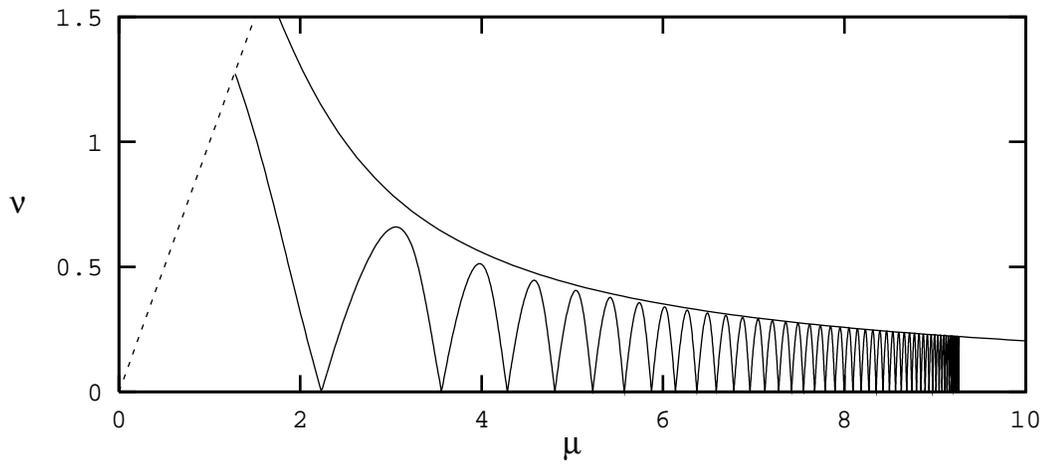

Figure 3: Three members of the periodic orbit family $10^n$ with (a) $n = 25$, (b) $n = 50$, and (c) $n = 100$ in the fundamental domain $\mu \geq \nu \geq 0$ of the regularized $\mu, \nu$ coordinates.

The surface of constant $J_h$, $J_c$ form a two dimensional torus in the 4–dimensional phase space. The total action of a trajectory on a torus after the time $T_c = 2\pi/\omega_c$, i.e. after one oscillation in $z$ – direction, is

$$J(\alpha, E) = J_c(\alpha, E) + \alpha J_h(\alpha, E), \tag{15}$$

where the winding number $\alpha = \omega_h/\omega_c$ is given as

$$\alpha = 2\omega J_c^3. \tag{16}$$

Inserting (12), (13) and (16) in (15), we can write the total action as function of the energy $E$ and $\alpha$,

$$J(\alpha, E) = \frac{\alpha}{2\omega} \left( E + \frac{3}{2} \left( \frac{2\omega}{\alpha} \right)^{2/3} \right). \tag{17}$$

In terms of the scaled energy $\epsilon = E\gamma^{-2/3}$ and $\gamma = 2\omega$, we obtain

$$J(\alpha, \epsilon) = \alpha \gamma^{-1/3} \left( \epsilon + \frac{3}{2} \alpha^{-2/3} \right). \tag{18}$$

The symbol '0' corresponds to one oscillation perpendicular to the field axis in the $\rho$ coordinates or half an oscillation in the original $(x, y)$ coordinates. A trajectory with $n$ consecutive zeros in the symbol code has an approximate winding number

$$\alpha(n) \approx n + \alpha_0 \quad \text{for} \quad n \gg 1. \tag{19}$$

Contributions due to the non – integrable part of the dynamics are collected here in the shift term $\alpha_0$, which approaches a constant in the limit $n \to \infty$ for periodic orbits in the same family. Inserting (19) in (18), the asymptotic $n$ dependent behavior of the (un–scaled) action of a channel trajectory can be written as

$$S(n) = 2\pi \gamma^{-1/3} \left[ n \left( \epsilon + \frac{3}{2} n^{-2/3} \right) + S_0 + S_{2/3} n^{-2/3} + \mathcal{O}(n^{-1}) \right]. \tag{20}$$

The coefficients $S_0, S_{2/3}, \ldots$ depend on $\epsilon$ and explicitly on the head string $G$ or on the past and future of the orbit before entering and after leaving the regular phase. Fig. 4 (a) shows results for the periodic orbit families $10^n$ and $20^n$ up to $n = 100$. The constant family dependent term in (20) is $S_0 = -0.6483$ for the $10^n$ family and $S_0 = -0.1440$ for the $20^n$ family, (see also Table 1).

The asymptotic behavior of the stability exponents for periodic orbits with long sequences of symbols '0' can be extracted from the trace of the stability– or Jacobi–matrix of the classical flow,

$$\mathbf{M}(t) = \frac{\partial(q(t), p(t))}{\partial(q_0, p_0)}. \tag{21}$$

This matrix describes the evolution of the classical dynamics in the neighborhood of a given trajectory (q(t),p(t)) in linear approximation.

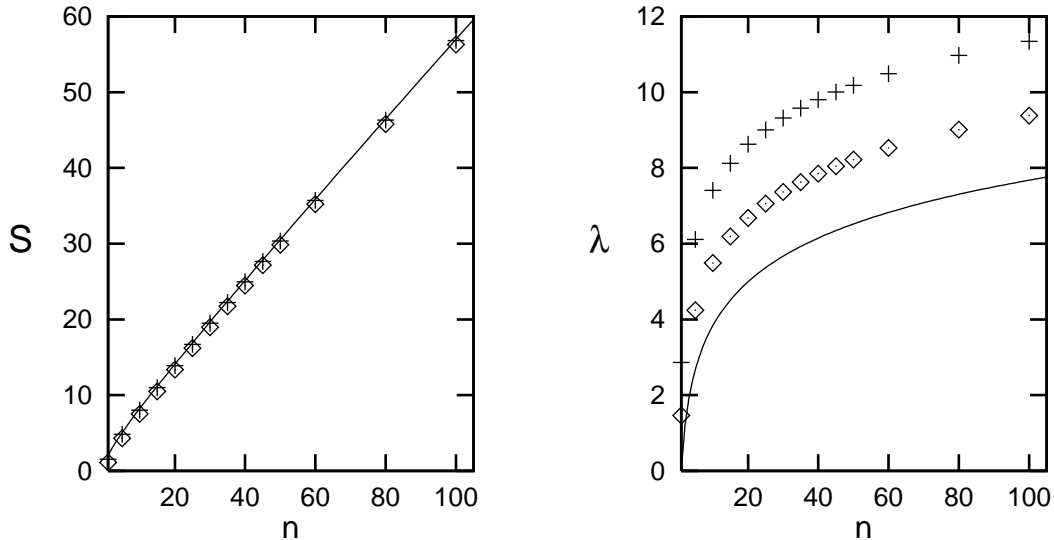

Figure 4: The actions (a) and stability exponents (b) of the periodic orbit families $10^n$ ($\diamond$) and $20^n$ ($+$) plotted with respect to the number $n$ of zeros in the symbol code (for $\epsilon = 0.5$); the full lines correspond to the asymptotic expressions $S_n \sim n(\epsilon + 3/2 n^{-2/3})$ and $\lambda_n = 5/3 \log(n)$, respectively.

The asymptotic behavior of $\mathrm{Tr}\mathbf{M}$ for a large number $n$ of symbols '0' is derived in Appendix C and we state only the final result here,

$$\mathrm{Tr}\mathbf{M} = k_1 n^{5/3} + k_2 n + \mathcal{O}(n^{2/3}). \tag{22}$$

The terms $k_1, k_2$ depend on $\epsilon$ and on the past and future of the periodic orbit before entering the regular region. The largest eigenvalue of the stability matrix grows approximately like the trace of $\mathbf{M}$. The stability exponent of a periodic orbit, i.e. the logarithm of the largest eigenvalue of the stability matrix, scales thus asymptotically like

$$\lambda(n) = \frac{5}{3} \log n + l_0 + l_{2/3} n^{-\frac{2}{3}} + \mathcal{O}(n^{-1}). \tag{23}$$

The logarithmic behavior and the factor $5/3$ is universal, i.e. independent of the scaled energy $\epsilon$ and of the head string $G$. In Fig. 4 (b), we compare periodic orbit data of the families $10^n$ and $20^n$ with the leading term $5/3 \log n$. The curves deviate not more than a constant in the large $n$ limit. This constant is again family dependent and we obtain $l_0 = 1.696$ for the $10^n$ – family and $l_0 = 3.675$ for the $20^n$–family (see Table 1). This is the slow logarithmic increase in the stability exponents, which reflects the regularity of the dynamic in the separable limit of the potential. The DKP shows the typical behavior of intermittency, i.e. the the Lyapunov exponent $\overline{\lambda} = \lambda(n)/T(n) \to 0$ for $n \to \infty$.

The discussion above is strictly valid only for scaled energies $\epsilon$ larger than the critical value $\epsilon_c$. The symbolic dynamics becomes pruned for $\epsilon < \epsilon_c$ and long strings of '0's tend to be forbidden below the critical value. The actual pruning rules depend on the past and future of the '0' code strings. The maximal possible length of the '0' string in a periodic orbit family, which becomes pruned, decreases with $\epsilon$. The $10^n$ family is an example showing this behavior. There are also families, as e.g. the $20^n$ family, where the limit $n \to \infty$ exists for all $\epsilon > 0$. A detailed analysis of these pruning rules will be presented elsewhere [36]. Note, however, that periodic orbit families with an infinitely long '0' tail allowed will have the same asymptotic behavior as discussed above. In the semiclassical analysis in section 3, we choose $\epsilon = 0.5$ to avoid problems due to the incomplete symbolic dynamics.

## 2 The Quantum Spectrum

The starting point of our quantum calculations is the Hamiltonian of a hydrogen atom in a uniform magnetic field, where we neglect relativistic corrections and effects of the finite nuclear mass recently discussed in [40, 41]. These effects occur on extremely long time scales, and correspondingly narrow energy spacings, which are accessible neither to present laboratory experiments nor to our numerical quantum calculations, and are also ignored in the classical and semiclassical calculations. The quantum Hamiltonian reads in atomic units and setting $\gamma = B/B_0$

$$H = \frac{1}{2}\mathbf{p}^2 - \frac{1}{r} + \frac{1}{2}\gamma L_z + \frac{1}{8}\gamma^2 \rho^2, \tag{24}$$

in analogy with its classical counterpart (2). Exact quantum numbers of the Hamiltonian (24) are the parity with respect to the $(z=0)$ – plane, $\pi_z$, and the magnetic quantum number, $m$, but the system remains non separable in $(\rho, z)$ coordinates. Eigenvalues and eigenstates of Schrödinger's equation can be calculated in each $m^\pi$ subspace separately.

Introducing dilated semiparabolic coordinates similar to (6)

$$\mu = \frac{1}{b}\sqrt{r+z} \, ; \quad \nu = \frac{1}{b}\sqrt{r-z}, \tag{25}$$

where $b$ is a free length scale parameter, Schrödinger's equation is transformed to

$$\left[ \triangle_\mu + \triangle_\nu + 2Eb^4(\mu^2 + \nu^2) - \frac{1}{4}b^8\gamma^2\,\mu^2\nu^2(\mu^2 + \nu^2) + 4b^2 \right] \Psi = 0 \tag{26}$$

where $\triangle_\mu$ and $\triangle_\nu$ are the Laplace operators for the radial coordinate in two dimensional space $(\mu, \varphi)$ and $(\nu, \varphi)$ with constant angular momentum $L = m$, i.e.

$$\triangle_\rho = \frac{1}{\rho}\frac{\partial}{\partial \rho}(\rho \frac{\partial}{\partial \rho}) - \frac{m^2}{\rho^2}, \quad (\rho = \mu \text{ or } \nu),$$

The paramagnetic term $\frac{1}{2}\gamma L_z$ at constant magnetic field strength results in a constant energy shift $E_\text{para} = m\gamma/2$, which is already omitted in (26).

The Schrödinger equation (26) depends on the two physical parameters $E$ and $\gamma$. Eigenvalues can now be calculated along various cuts in this plane, e.g. at constant magnetic field strength $\gamma$, along lines $E\gamma = $ const or along lines of constant scaled energy $\epsilon = E\gamma^{-2/3}$. For example at constant magnetic field strength we introduce

$$\lambda \equiv -(1 + 2Eb^4),$$

and obtain a generalized eigenvalue equation for the parameter $\lambda$

$$\left[H_0 - \frac{1}{4}b^8\gamma^2\,\mu^2\nu^2(\mu^2 + \nu^2) + 4b^2\right]\Psi = \lambda\left(\mu^2 + \nu^2\right)\Psi \qquad (27)$$

with

$$H_0 = (\triangle_\mu - \mu^2) + (\triangle_\nu - \nu^2). \qquad (28)$$

Due to scaling laws of the classical Hamiltonian (2), the structure of the classical dynamics depends only on the scaled energy $\epsilon = E\gamma^{-2/3}$. For comparison with semiclassical theories, it is therefore most convenient to evaluate the quantum spectra along lines $\epsilon$ constant. Choosing

$$s \equiv \gamma^{-1/3} \quad ; \quad b \equiv \beta s = \text{const}$$

with $\beta$ a new free parameter, (26) can be transformed into a generalized eigenvalue expression for $s$

$$\left[-2\epsilon(\mu^2 + \nu^2) + \frac{1}{4}\beta^8\,\mu^2\nu^2(\mu^2 + \nu^2) - 4\beta^2\right]\Psi = s^{-2}\left(\triangle_\mu + \triangle_\nu\right)\Psi \quad . \qquad (29)$$

Both Eqs. (27) and (29) can now be represented in matrix form as a generalized eigenvalue problem with sparse symmetric matrices in terms of the complete set of basis functions given by products of the eigenstates $|N_\rho m\rangle$ of the two-dimensional harmonic oscillator (28), viz.

$$|N_\mu N_\nu m\rangle = |N_\mu m\rangle \otimes |N_\nu m\rangle \quad ; \quad N_\mu, N_\nu = 0, 1, 2, \ldots \quad . \qquad (30)$$

Eigenvalues $\lambda = -(1 + 2b^4 E)$ or $s = \gamma^{-1/3}$ and the corresponding eigenstates are obtained by numerical diagonalization of matrices.

Because the basis states (30) are square integrable $\mathcal{L}_2$ functions scattering wave functions of open systems can not be expanded in this basis and the method is usually restricted to the calculation of *bound states* at energies below the ionisation threshold, i.e.

$$E < E_{\rm IP} = \frac{\gamma}{2}(|m| + 1)$$

or for constant scaled energy $\epsilon > 0$

$$s = \gamma^{-1/3} < \frac{|m| + 1}{2\epsilon}.$$

Note, that the spectrum of hydrogen in a magnetic field exhibits bound states at energies even above the classical threshold $E = 0$, i.e. when the classical motion is already unbound in $z$ direction. The continuum wave functions are (for $|z| \to \infty$) still quantized in the $\rho$ – coordinate and the quantum threshold is thus shifted by the zero point energy of the harmonic oscillator in the classical Landau channel. To account for *unbound resonances* in the continuum we adopted the complex-rotation method [42, 43], which is based on the replacement

$$\mathbf{r} \longrightarrow \mathbf{r}\, e^{i\theta}$$

in the Hamiltonian and in the wave functions. By this transformation, hidden resonances of the Hamiltonian in the continuum, associated with complex eigenvalues, are exposed, while the resonance wave functions can still be described by the $\mathcal{L}_2$ integrable basis functions (30), but with complex arguments. In our approach, the complex rotation by the angle $\theta$ is reflected in the replacement of $b$ and $\beta$ in Eqs. (27) and (29) with the complex dilatation parameter

$$b = |b|\, e^{i\frac{\theta}{2}} \quad \text{and} \quad \beta = |\beta|\, e^{i\frac{\theta}{2}},$$

respectively. Representing Eqs. (27) and (29) with complex rotated coordinates in matrix form, this results in a generalized eigenvalue problem for complex symmetric, non-Hermitian matrices. Numerically, the eigenvalue equations were solved by extending the Lanczos algorithm [44] to complex matrices.

If the scale parameter $b$ or $\beta$ and the complex rotation angle $\theta$ are appropriately chosen and a sufficiently high number of basis functions (30) is considered in the numerical diagonalization isolated bound states along the real axis as well as resonances in the lower complex half plane can be obtained to high numerical precision. The method fails close to the Landau thresholds

$$E_N = \gamma \left( N + \frac{|m|+1}{2} \right) \tag{31}$$

or, for constant scaled energy $\epsilon > 0$

$$s_N = \frac{1}{\epsilon} \left( N + \frac{|m|+1}{2} \right) \tag{32}$$

with $N = 0, 1, 2, \ldots$, the Landau quantum number. To each Landau channel belongs an infinite number of Rydberg like bound states ($N = 0$) or resonances ($N > 0$) converging to the threshold. Obviously such infinite series of states can not be obtained from a calculation with a finite basis size. The numerical convergence of resonances close to the thresholds is extremely slow if all basis functions up to a given oscillator quantum number, i.e.

$$N_\mu + N_\nu + |m| \leq N_{\max}$$

are considered in the calculation because the total basis size increases quadratically with growing $N_{\max}$. To improve the convergence we chose a different method to truncate the basis. In coordinate space states $|N_\mu N_\nu m>$ with $N_\mu \approx N_\nu$ are localized close to the $(z=0)$-plane. In this direction the classical motion is bound and there is no need to complete the basis with states of this type. On the other hand states with $N_\mu \ll N_\nu$ or $N_\nu \ll N_\mu$ are oriented along the $z$-axis, i.e. where the potential is open. These states are important for the representation of Rydberg like resonances reaching far into the Landau channels. We have now optimized our basis by taking into account only those states which cover the physically important region. With this choice the numerical efficiency in the calculation of Rydberg like resonances could be considerably improved.

Results for a quantization along constant $\epsilon$–lines will be presented in the next section in comparison with the semiclassical calculations.

## 3 Semiclassical quantization

In the following, we will present a method for a semiclassical quantization of hydrogen in a constant magnetic field for $L_z = 0$, i.e. with azimuthal quantum number $m = 0$. The quantization will be performed with respect to the magnetic field variable $s = \gamma^{-1/3}$ introduced in the last section along constant scaled energy lines $\epsilon = E\gamma^{-2/3} > \epsilon_c$.

Starting point of our semiclassical calculation is the zeroth order Gutzwiller–Voros zeta function [27], which is the leading term in a semiclassical approximation of the spectral determinant $\det(E - \hat{H})$. The zeta function can be derived from the periodic orbit trace formula [22, 45] and has for the scale invariant classical Hamiltonian (2) with fixed scaled energy $\epsilon$ the form

$$\zeta^{-1}(s) = \prod_p \left(1 - t_p(s)\right), \quad \text{with } t_p(s) = e^{is\tilde{S}_p - i\sigma_p \frac{\pi}{2} - \frac{\lambda_p}{2}}. \tag{33}$$

The product runs over single repeats of all periodic orbits of the system. The scaled action $\tilde{S}$ introduced in (5) is taken here along a periodic orbit. The (complex) variable $s$ is the scaling factor $\gamma^{-1/3}$ in units of $\hbar$. The winding number $\sigma$ corresponds to the number of turns of the stable or unstable invariant manifold around the periodic orbit [46]. The linear stability exponent $\lambda$ is defined as the logarithm of the largest eigenvalue of the stability matrix $\mathbf{M}$, Eq. (21). (We restrict ourself here to the purely hyperbolic case, i.e. we presume that all periodic orbits are unstable and $\epsilon > \epsilon_c$).

The semiclassical approximation to the quantum eigenvalues are given by the zeros of the zeta function. The magnetic field dependence enters the semiclassical zeta function for fixed $\epsilon$ only through the scaling factor $s = \gamma^{-1/3}$. The winding number and the stability exponent are independent of $\gamma$ due to the scale invariance of the classical Hamiltonian. A quantization with respect to the magnetic field variable $s$ uses the classical information in the most optimal way as the periodic orbits have to be calculated only once for each $\epsilon$.

The symmetry reduced zeta functions corresponding to the sub–spectra with fixed parity $\pi_z = \pm 1$ are obtained by restricting the classical dynamics to the fundamental domain, see Appendix A. The symmetric and anti–symmetric zeta functions differ in phase factors, which are due to the different boundary conditions as explained in detail in Appendix A. These phase factors are included here in the Maslov index $\sigma$, which can be read off directly from the three letter symbolic dynamics introduced in section 1.1. We obtain [31]

$$\sigma_p = 2n_p + n_2 + (1 - \pi_z)n_1, \tag{34}$$

where $n_p$ denotes the length of the symbol string corresponding to the periodic orbit, and $n_1$, $n_2$ count the number of symbols '1', '2' in the code. Note, that the periodic orbit 2 running along the symmetry line has to be excluded in (33) for $\pi_z = -1$.

The product representation of the zeta function (33) is absolute convergent only for $Im(s) > h_a > 0$ [47]. For systems with a well defined symbolic dynamics a so called *cycle expansion* of products like (33) has been proposed [28, 29] to provide a representation with larger analyticity domain. The product is expanded by multiplying out the single factors and regrouping the terms in such a way that maximal cancellations occur. The ordering scheme makes intensive use of the self–similar structure of the dynamics reproduced by the symbolic description of the flow and groups together orbits and pseudo–orbits into contributions of increasing total symbol length, the so called curvature terms. For a symbolic dynamics with a three letter alphabet $\{0,1,2\}$, this has the form [29]

$$\zeta^{-1} = \prod_p \left(1 - t_p\right) = 1 - \sum_{n=1}^{\infty} c_n \qquad (35)$$

with

$$\begin{aligned}
c_1 &= [t_0 + t_1 + t_2] \qquad \text{(fundamental term)} \\
c_2 &= [(t_{01} - t_0 t_1) + (t_{02} - t_0 t_2) + (t_{12} - t_1 t_2)] \\
c_3 &= [(t_{001} - t_0 t_{01}) + (t_{011} - t_{01} t_1) + (t_{002} - t_0 t_{02}) \\
&\quad + (t_{022} - t_{02} t_2) + (t_{112} - t_1 t_{12}) + (t_{122} - t_{12} t_2) \\
&\quad + (t_{021} - t_{02} t_1) + (t_{012} - t_0 t_{12} - t_{01} t_2 + t_0 t_1 t_2)] \\
c_4 &= \ldots \quad .
\end{aligned}$$

Note that each term in the curvature contributions $c_n$ is accompanied by a 'shadowing' term differing in sign. The only exception is the leading term $c_1$, which is also called the fundamental term [29] in the expansion.

The method has been applied successfully for variety of strictly hyperbolic systems [48, 29, 49, 50] and yields an analytic continuation of the zeta function in a strip $h_a \geq Im(s) > h_c$, $h_c < 0$ with exponentially decreasing curvature terms $c_n$ in this regime. Analytic semiclassical expressions below the critical value $h_c$ can be obtained by modifications of the Gutzwiller–Voros zeta function itself [51], which will not be discussed here. Hyperbolic systems are characterized by Lyapunov exponents $\overline{\lambda}_p = \frac{\lambda_p}{S_p}$ of periodic orbits strictly bounded away from 0. The DKP, however, is an intermittent system even for scaled energies $\epsilon$ above the critical value $\epsilon_c$ and we obtain $\overline{\lambda} \sim \frac{\log n}{n} \to 0$ for periodic orbits with an increasing '0'– tail in the symbol code. This behavior has strong implications on the curvature expansion and a possible analytic continuation of the zeta function, which will be discussed in the next paragraph.

## 3.1 Cycle expansion with infinite symbolic dynamics

The dynamics of the DKP is strongly influenced by the regular behavior in the asymptotically separable channel as discussed in detail in section 1.2. Periodic orbits with equal symbol length contribute very differently to the zeta function depending on the time spend far from the nucleus. The weight $t_p$ of channel trajectories drop off algebraically with the period, whereas contributions from periodic orbits localized in the center region decrease exponentially in amplitude with increasing symbol length. As a consequence, shadowing properties in a cycle expansion (35) between orbits with equal symbol code length are poor. Furthermore, the regular orbits dominate the curvature terms $c_n$ with increasing symbol length $n$ giving rise to an algebraic decay in the curvature expansion

$$c_n \approx n^{-5/6} e^{2\pi i s \epsilon n}$$

for real $s$. As a consequence, an expansion of the zeta function grouping together single orbit terms like in (35) diverges for $Im(s) < 0$, i.e. exactly in the region, where the quantum resonances are expected.

Controlling the divergences introduced through the coexistence of regular and chaotic motion is one of the main challenges in classical and semiclassical periodic orbit theory [52]. In the following, we present a method which overcomes the problem of including intermittency originating from one marginally stable orbit in the cycle expansion of a zeta function. We point out, however, that the distribution of marginally orbits in a generic system with mixed regular and chaotic motion, as e.g. in the classical DKP for $\epsilon < \epsilon_c$, may itself have a complicated structure.

The marginally stable behavior for $|z| \to \infty$ enters the symbolic dynamics through the symbol '0', which corresponds to half an oscillation perpendicular to the magnetic field axis. The symbols '1' and '2' are always related to motion in the near core region. The symbol '0' can occur everywhere in the regular channel arbitrary far from the origin. A single '0' in the symbol code contributes differently to the action or stability exponent of a periodic orbit depending on how far from the nucleus the corresponding oscillation takes place. The classical dynamics is thus not reflected by the symbol '0' alone, but by the number of '0's occurring successively in a row.

Consequently, we have to change our 3 – symbol alphabet to a double infinite alphabet in the following way; substrings starting with a '1' or '2' and followed by $n-1$ symbols '0' are transformed according to the rule

$$10^{n-1} \to n^1, \quad 20^{n-1} \to n^2; \quad n = 1, 2, \ldots, \infty. \tag{36}$$

The index $n$ corresponds to the total length of the substring in the ternary alphabet. The symbol code length of a periodic orbit in the infinite alphabet is given by the number of symbols differing from '0' in the ternary code. All periodic orbits in the families $10^{n-1}, 20^{n-1}$ correspond now to orbits $n^1, n^2$ with symbol length one in the new alphabet. The orbits with symbol length two can be written as

$$(mn)^{11}, (mn)^{12}, (mn)^{22} \quad m, n = 1, 2, \ldots \infty.$$

The upper index denotes here the starting symbols '1' or '2' of the different '0' substrings, the lower index corresponds to the length of the substrings. We realize, that the new symbolic dynamics has a matrix structure, e.g. $(mn)^{11}$ corresponds to all codes $10^{m-1}10^{n-1}$, $(mn)^{12}$ includes all strings $10^{m-1}20^{n-1}$ and so on. Table 1 lists the possible symbols up to length three in the new alphabet together with some periodic orbit families. A periodic orbit family is defined in analogy to section 1.2 as all periodic orbits which differ only in one number in the lower index. Note that a cyclic shift both in the upper and lower index denotes the same periodic orbits. The symbol matrices $(mn)^{12}$ and $(mn)^{21}$ are thus equivalent, i.e. $(mn)^{12} = (nm)^{21}$ and the matrices $(mn)^{11}$, $(mn)^{22}$ are symmetric.

| ∞ – alphabet | ternary alphabet | | | | | |
|---|---|---|---|---|---|---|
|  | $n=1$ | $n=2$ | $n=3$ | $n=4$ | $S_0$ | $l_0$ |
| $n^1$ | 1 | 10 | 100 | 1000 | $-0.648$ | 1.696 |
| $n^2$ | 2 | 20 | 200 | 2000 | $-0.144$ | 3.675 |
| $(mn)^{11}$ | | | | | | |
| $(1n)^{11}$ | 11 | 110 | 1100 | 11000 | $-0.243$ | 3.126 |
| $(2n)^{11}$ | 101 | 0101 | 10100 | 101000 | 0.342 | 4.220 |
| $(3n)^{11}$ | 1001 | 10010 | 100100 | 1001000 | 0.692 | 5.000 |
| $(mn)^{12}$ | | | | | | |
| $(1n)^{21}$ | 21 | 210 | 2100 | 21000 | $-0.235$ | 3.011 |
| $(2n)^{21}$ | 201 | 2010 | 20100 | 201000 | 0.628 | 5.412 |
| $(3n)^{21}$ | 2001 | 20010 | 200100 | 2001000 | 1.053 | 6.460 |
| $(1n)^{12}$ | 12 | 120 | 1200 | 12000 | 0.628 | 5.412 |
| $(2n)^{12}$ | 102 | 1020 | 10200 | 102000 | 1.052 | 6.460 |
| $(3n)^{12}$ | 1002 | 10020 | 100200 | 1002000 | 1.345 | 7.161 |
| $(mn)^{22}$ | | | | | | |
| $(1n)^{22}$ | 22 | 220 | 2200 | 22000 | 0.838 | 6.287 |
| $(2n)^{22}$ | 202 | 0202 | 20200 | 202000 | 1.388 | 7.767 |
| $(3n)^{22}$ | 2002 | 20020 | 200200 | 2002000 | 1.725 | 8.692 |
| $(kmn)^{111}$ | | | | | | |
| $(11n)^{111}$ | 111 | 1110 | 11100 | 111000 | 0.438 | 4.562 |
| $(kmn)^{112}$ | | | | | | |
| $(11n)^{112}$ | 112 | 1120 | 11200 | 112000 | 1.225 | 6.928 |
| $(11n)^{211}$ | 211 | 2110 | 21100 | 211000 | 0.605 | 5.297 |
| $(11n)^{121}$ | 121 | 1210 | 12100 | 121000 | 0.605 | 5.297 |
| $(kmn)^{122}$ | | | | | | |
| $(11n)^{122}$ | 122 | 1220 | 12200 | 122000 | 1.644 | 8.220 |
| $(11n)^{212}$ | 212 | 2120 | 21200 | 212000 | 1.348 | 7.582 |
| $(11n)^{221}$ | 221 | 2210 | 22100 | 221000 | 0.844 | 6.321 |
| $(kmn)^{122}$ | | | | | | |
| $(11n)^{222}$ | 222 | 2220 | 22200 | 222000 | 1.868 | 9.169 |

Table 1: Infinite alphabet compared with the ternary alphabet for the shortest periodic orbit families; the last columns denote the leading coefficients in the expansions (20), (23) for $\epsilon = 0.5$.

The cycle expansion in our new double–infinite alphabet has now the form

$$\zeta^{-1} = \prod_p (1 - t_p) = 1 - \sum_{n=1}^{\infty} c_n \tag{37}$$

with

$$\begin{aligned}
c_1 &= \sum_{n=1}^{\infty} \left[ t_n^1 + t_n^2 \right] \qquad \text{(fundamental term)} \\
c_2 &= \sum_{m=1}^{\infty} \sum_{n=1}^{\infty} \left[ \frac{1}{2}(t_{mn}^{11} - t_m^1 t_n^1) + (t_{mn}^{12} - t_m^1 t_n^2) + \frac{1}{2}(t_{mn}^{22} - t_m^2 t_n^2) \right] \\
c_3 &= \sum_{k=1}^{\infty} \sum_{m=1}^{\infty} \sum_{n=1}^{\infty} \left[ \left( \frac{1}{3} t_{kmn}^{111} - \frac{1}{2} t_{km}^{11} t_n^1 + \frac{1}{6} t_k^1 t_m^1 t_n^1 \right) \right. \\
&\qquad + \left( t_{kmn}^{112} - t_k^1 t_{mn}^{12} - \frac{1}{2} t_{km}^{11} t_n^2 + \frac{1}{2} t_k^1 t_m^1 t_n^2 \right) \\
&\qquad + \left( t_{kmn}^{122} - t_{km}^{12} t_n^2 - \frac{1}{2} t_k^1 t_{mn}^{22} + \frac{1}{2} t_k^1 t_m^2 t_n^2 \right) \\
&\qquad \left. + \left( \frac{1}{3} t_{kmn}^{222} - \frac{1}{2} t_{km}^{22} t_n^2 + \frac{1}{6} t_k^2 t_m^2 t_n^2 \right) \right] \\
c_4 &= \sum_{l=1}^{\infty} \sum_{k=1}^{\infty} \sum_{m=1}^{\infty} \sum_{n=1}^{\infty} \ldots
\end{aligned}$$

The weights $t_p$ are written here as the elements of infinite dimensional tensors of rank $n_c$, where $n_c$ is the length of the symbol string in the infinite alphabet or equivalent the length of the upper index. The curvature terms $c_n$ in the cycle expansion (37) are now sums over all the elements of the tensors of rank $n$ including outer products of tensors with rank $n' < n$. The prefactors are introduced to avoid double counting of periodic orbits. Note that the sum over the prefactors alone (including the sign) adds up to zero in each curvature term.

We focus first on the leading or fundamental term $c_1$ in the cycle expansion, which is a sum over all members of the periodic orbit families $n^1$, $n^2$. In order to study the behavior of a single periodic orbit family sum, we introduce the notation

$$\hat{t}^a(s) = \sum_{n=1}^{\infty} t_n^a = \sum_{n=1}^{\infty} e^{2\pi i (s\tilde{S}^a(n) - n(N+1/2)) - \lambda^a(n)/2}, \tag{38}$$

with an upper index $a = \{1, 2\}$. An additional phase $2\pi N n$ is added (with $N$ integer), which do not affect the sum. Its meaning will become clear in what follows. (Note there is another phases $\pi/2$ in the case of $a = \,$'2', see also Eq. (34).)
Using the asymptotic results (20), (23) from section 1.2, we write

$$\tilde{S}^a(n) = n\left(\epsilon + \frac{3}{2}n^{-\frac{2}{3}}\right) + S_0^a + S_{2/3}^a n^{-2/3} + \sum_{i=1}^{6} S_i^a n^{-i}; \qquad (39)$$

$$\lambda^a(n) = \frac{5}{3}\log n + l_0^a + l_{2/3}^a n^{-2/3} + \sum_{i=1}^{6} l_i^a n^{-i}, \qquad (40)$$

and $\tilde{S}$ is written here in units of $2\pi$. The family dependent coefficients $S_0, S_{2/3}, \ldots, l_0, l_{2/3}, \ldots$ are obtained from a polynom fit on periodic orbit data up to a symbol length $n = 500$ in the ternary alphabet. The leading coefficients $S_0^{(a)}$ and $l_0^{(a)}$ for $\epsilon = 0.5$ are listed in Table 1.

The sums $\hat{t}^a$ diverge for $Im(s) < 0$ due to the logarithmic behavior in the stability exponent in (40). This is the same kind of divergence discussed earlier, and it appears in all other curvature terms. An analytic continuation of the sums (38) for negative imaginary $s$ values is provided by the following technique [32]; first we transfer the periodic orbit sum into a sum over integrals, using the identity

$$\sum_{n=-\infty}^{\infty} \delta(x-n) = \sum_{r=-\infty}^{\infty} \exp(2\pi i r x) \qquad (41)$$

obtained by Poisson summation. The periodic orbit sums $\hat{t}^a$ can thus be written as

$$\hat{t}^a(s) = \frac{1}{2}t_1^a(s) + \sum_{r=-\infty}^{\infty} \int_0^{\infty} dx\, e^{2\pi i r x} t^a(s, x+1). \qquad (42)$$

The continuous variable $x$ in $t^a(s,x)$ corresponds here to the discrete symbol length $n$ in (38). We now fix the index $N$ in (38) by the condition

$$Re(s)\,\epsilon + \frac{1}{2} > N > Re(s)\,\epsilon - \frac{1}{2},$$

i.e. we change $N$ exactly at the Landau thresholds (32) $Re(s) = (N+1/2)/\epsilon$. The $r = 0$ integral can then be evaluated by rotating the line of integration onto the negative imaginary axis which corresponds to a transformation $x \to -\mathrm{i}x$. The integrand is now exponentially decreasing for all imaginary parts of $s$ and $Re(s) \neq (N+1/2)/\epsilon$. The integral is thus convergent for $Im(s) < 0$, but we pick up an essential singularity at the Landau thresholds. By the same technique, we find the analytic continuation for the other $r$ – integrals. The real axis has to be rotated according to $x \to \pm\mathrm{i}x$ and the $\pm$–sign refers to the sign of $r$. We end up with the compact expression

$$\hat{t}^a(s) = \frac{1}{2}t_1^a(s) - \mathrm{i}\int_0^{\infty} dx\, t^a(s, 1-\mathrm{i}x) \qquad (43)$$
$$+ \mathrm{i}\int_0^{\infty} dx\, \frac{e^{-2\pi x}}{1-e^{-2\pi x}}\left[t^a(s, 1+\mathrm{i}x) - t^a(s, 1-\mathrm{i}x)\right].$$

The sum over $r$ for $r \neq 0$ is nothing but a geometric series which gives rise to the second integral in (43). The transformation from the original sum to the two integrals in (43) is indeed an equality for $Im(s) \geq 0$, but provides an analytic continuation for $Im(s) < 0$. The second curvature terms can be worked out in the same spirit. The leading contributions to the action and stability exponents are

$$\tilde{S}^a(n) = n\left(\epsilon + \frac{3}{2}n^{-\frac{2}{3}}\right) + m\left(\epsilon + \frac{3}{2}m^{-\frac{2}{3}}\right) + S_0^a + \mathcal{O}(n^{-\frac{2}{3}} + m^{-\frac{2}{3}}) \qquad (44)$$

$$\lambda^a(n) = \frac{5}{3}\log(nm) + l_0^a + \mathcal{O}(n^{-\frac{2}{3}} + m^{-\frac{2}{3}})$$

and 'a' corresponds to '11', '12' or '22', here. The analytic continuation of the double sum $c_2$ in (37) leads to a double integral, the higher curvature terms give rise to higher dimensional integrals, respectively.

Before presenting results for the full cycle expansion, we will study the leading term $c_1$ in detail. We will show, that this term is indeed fundamental in the sense, that it contains already all the information about the gross structure of the spectrum.

## 3.2 The QEBK – Quantization

The dominant contribution to the periodic orbit family sums (38) is contained in the $r = 0$ integral in (42). The other $r$ – terms give a contribution suppress by the factor $e^{-2\pi x}$, as can be seen from Eq. (43). To obtain the dominant $s$–dependent behavior of the periodic orbit family sums, we approximate the integral

$$I^a(s) = \int_1^\infty dx\, e^{2\pi i\left[s\tilde{S}^a(x) - (N+1/2)x\right] - \lambda^a(x)/2} \qquad (45)$$

by stationary phase. This leads to the phase condition

$$\frac{\partial}{\partial x}\left(u\tilde{S}^a(x) - (N + \frac{1}{2})x\right) = 0, \qquad (46)$$

where we set $s = u + iv$. The solutions of (46) are in leading order (up to corrections $\mathcal{O}(x^{-5/3})$) given by

$$x_0(u) = \left[\frac{u}{2(N + \frac{1}{2} - u\epsilon)}\right]^{3/2}. \qquad (47)$$

Note that $x_0$ tends to infinity at the Landau thresholds $u = (N + \frac{1}{2})/\epsilon$. We end up with the approximation to the integral

$$I^a(s) \approx \left|u\frac{\partial^2 \tilde{S}^a}{\partial x^2}(x_0)\right|^{-1/2} e^{2\pi i\left[s\tilde{S}^a(x_0) - (N+1/2)x_0 - 1/8\right] - \lambda^a(x_0)/2} \qquad (48)$$

$$\stackrel{x_0 \gg 1}{\approx} \sqrt{3}\, e^{-l_0^a/2} u^{-1/2} e^{2\pi i\left[sx_0(\epsilon + \frac{3}{2}x_0^{-2/3}) + sS_0^a - (N+1/2)x_0 - 1/8\right]} + \mathcal{O}(x_0^{-2/3})$$

and $x_0(u)$ is the solution of Eq. (46). Note that the exponent 5/3 obtained from the leading behavior of the stability exponents in (40) exactly cancels the singularity introduced through the second derivatives of the actions at the Landau thresholds.

The additional term 1/8 in the exponent is due to the stationary phase approximation.

In a next step we approximate the term $c_1$ in the cycle expansion (37) by the dominating periodic orbit family only, i.e. by the family with the smallest coefficient $l_0$ in (40). For $\epsilon = 0.5$, this is the '1' – family with $l_0 = 1.696\ldots$, see Table 1. This leads to the approximate quantization condition,

$$\zeta^{-1}(s) \approx 1 - \hat{t}^1(s) = 0. \tag{49}$$

Inserting the stationary phase expression (48) in (49), we can write down a quantization condition for the real parts of $s$ directly,

$$u\,\tilde{S}^1(x_0(u)) - (N + \frac{1}{2})\,x_0(u) = M + \frac{1}{8} \tag{50}$$

with $M$ integer. The quantization conditions (46) together with (50) can be interpreted in terms of the semiclassical Einstein–Brillouin–Keller (EBK) quantization of integrable systems, (see [22] for an overview). The Hamiltonian of a classical integrable systems with $f$ degrees of freedom can be written in the form $H = H(\vec{J})$, and $J_i$, $i = 1,\ldots,f$ denotes the conserved momenta (or actions). An example of such an Hamiltonian is given in (11). The EBK treatment leads to the simple quantization condition $J_i = n_i + \frac{\nu_i}{4}$ and the energy eigenvalues are obtained from inserting the quantized actions in the classical Hamiltonian. The integer numbers $\nu_i$ are called the Maslov indices and correspond to the number of caustics along a path $\varphi(t) = \omega_i t$ on the torus.

For the integrable Hamiltonian (11), we can express the action after one oscillation along the magnetic field direction by the equation (15), i.e.

$$J(\alpha, E) = J_c(\alpha, E) + \alpha J_h(\alpha, E), \tag{51}$$

The variable $\alpha$ corresponds to the winding number in Eq. (16). Due to the condition $\frac{\partial J_c}{\partial \alpha} = -\alpha \frac{\partial J_h}{\partial \alpha}$ valid for fixed energy, we can write the additional condition

$$\frac{\partial J}{\partial \alpha} = J_h \tag{52}$$

for $E$ constant. Equations (51) and (52) are indeed equivalent to (50) and (46), if we make the natural identification $J = u\tilde{S}^1$ and $\alpha = x$ and choose the quantization conditions

$$J_c(s,\epsilon) = M + \frac{1}{8} \quad , J_h(s,\epsilon) = N + \frac{1}{2}. \tag{53}$$

The harmonic oscillator is quantized with the usual term 1/2 corresponding to an integer Maslov index 2. We are lead to a non integer Maslov phase 1/2 or a term 1/8 in the quantization rule for the action $J_c$ in the Coulomb direction. (Similar results have been found in Helium, see Ref. [32]). A full winding in the $J_c$ coordinate implies a "visit" of the near core region, where the torus approximation of the full dynamics breaks down. This fact may motivate the unusual Maslov term, a clear understanding within the EBK – theory is, however, still missing. Equations similar to (51–53) have been applied to quantize stable islands in the classical phase space beyond the harmonic oscillator approximation assuming near-integrability of dynamics in the neighborhood of the stable fix-point [53, 54]. Note, that the quantization rules are derived here directly from Gutzwiller's periodic orbit formula and provide a quantization both in the real and imaginary part of the $s$ – variable.

An effective quantization scheme in terms of equations (50) and (46) is obtained by first quantizing the winding number $x = x_{N,M}$ according to

$$\tilde{S}(x)(N + \frac{1}{2}) - \frac{\partial \tilde{S}}{\partial x}\left[(N + \frac{1}{2})x + (M + \frac{1}{8})\right] = 0. \qquad (54)$$

The eigenvalues $u_{N,M}$ are finally given by the formula

$$u_{N,M} = \frac{(N + 1/2)\, x_{N,M} + (M + \frac{1}{8})}{\tilde{S}(x_{N,M})} \qquad (55)$$

Note that there is no free parameter. The action $\tilde{S}(x)$ is determined by Eq. (39), and the family dependent coefficients in the expansion are obtained from the periodic orbit data.

Asymptotic solutions of the Eq. (54), (55) near the Landau thresholds can be given in analytic form,

$$u_{N,M}^{-1} = \frac{\epsilon}{N + \frac{1}{2}} + \frac{1}{2(N + \frac{1}{2})}\left[\epsilon\frac{M + 1/8}{N + 1/2} - S_0^1(\epsilon) + \frac{1}{2\epsilon^2}\frac{N + 1/2}{M + 1/8}\right]^{-2}, \; M \gg 1, (56)$$

where the parameter $S_0^1(\epsilon)$ denotes the leading coefficient of the expansion of the action Eq. (39) of the dominating periodic orbit family, (which is the $n^1$ family for $\epsilon = 0.5$).

In Tables 2–4, we list the energy eigenvalues obtained from our EBK – like quantization (54) and (55), which we call quasi–EBK (QEBK) quantization from now on [32]. They are compared with the real parts of quantum eigenvalues calculated from Schrödinger's equation directly as described in section 2. Our simple quasi – integrable approach can reproduce the quantum spectra already surprisingly well. The QEBK results deviate in large parts of the spectrum not more than 10% of the mean level spacing and the bound states in the series below the first Landau threshold ($Re(s) < 1/2\epsilon = 1$) are completely reproduced within an error of $\delta < 4\%$. This makes it possible to assign approximate quantum numbers $(N, M)$ to individual states in the spectrum. Note, that there are more quantum resonances than approximate quantum numbers obtained from our quasi separable ansatz. The eigenstates at $Re(s) = 2.44, 3.73$ and $4.64$ in Table 3, 4 cannot be reproduced by the QEBK – approach. These states have thus no equivalent in the separable problem (11), and are an effect of the coupling region in the full potential $V(\rho, z)$ in (2). In other words, the phase space volume of the non–separable problem is (for finite $|z|$) larger than the corresponding volume in the separable approximation, which causes additional resonances. Their number can be estimated by the difference in the phase space volume and is expected to increase like $s^2$ in the spectrum.

|   |   | Quantum Mechanics | | QEBK | | Cycle Expansion | | |
|---|---|---|---|---|---|---|---|---|
| $N$ | $M$ | Re $s$ | Im $s$ | Re $s$ | $\delta$ | Re $s$ | Im $s$ | $\delta$ |
| 0 | 0 | 0.53581 | 0.00000 | 0.54072 | 1.07 | 0.50685 | 0.01972 | 6.06 |
| 0 | 1 | 0.81404 | 0.00000 | 0.81453 | 0.34 | 0.80786 | 0.00480 | 4.17 |
| 0 | 2 | 0.90117 | 0.00000 | 0.90212 | 1.61 | 0.89841 | 0.00225 | 4.58 |
| 0 | 3 | 0.94017 | 0.00000 | 0.94085 | 2.43 | 0.93879 | 0.00120 | 4.79 |
| 0 | 4 | 0.96039 | 0.00000 | 0.96084 | 2.94 | 0.95961 | 0.00069 | 4.97 |
| 0 | 5 | 0.97203 | 0.00000 | 0.97233 | 3.20 | 0.97155 | 0.00043 | 5.16 |
| 0 | 6 | 0.9795 | 0.0000 | 0.97947 | 0.56 | 0.97896 | 0.00028 | 9.23 |
| 0 | $\infty$ | 1.0 | 0.0 | 1.0 | | 1.0 | 0.0 | |

Table 2: The spectrum of hydrogen in a constant magnetic field for $m^\pi = 0^+$ and $\epsilon = 0.5$; the 3$^{\rm rd}$ and 4$^{\rm th}$ columns contain the quantum results for $s = \gamma^{-1/3}$ calculated by the complex rotation method, see section 2; the approximate quantum numbers $N$ and $M$ are obtain from the QEBK results in column 5. The zeros of the cycle expanded zeta function are presented in columns 7 and 8; the deviation $\delta$ of the real parts of the semiclassical and quantum results (in percent of the mean level spacing in each Rydberg series) is given in columns 6 and 9.

The approximate quantum numbers also help to clarify the structure of the spectrum. The Rydberg series converging to the different Landau thresholds $Re(s) = (N+1/2)/\epsilon$ overlap for $N \geq 1$, the bound state series $N = 0$ is the only one remaining unperturbed. We identify 3 perturber states in the $N = 1$ series and 9 perturber in the $N = 2$ series. The number of perturber states increase rapidly with increasing $N$. The perturber states interfere with neighboring Rydberg states causing level repulsion both in the real and imaginary part of the eigenvalues. The interference effects are a result of the non–separability of the quantum Hamiltonian and can not be resolved within the QEBK approach. As a consequence, the error in the QEBK–eigenvalues is maximal in the neighborhood of a perturber and a unique assignment of quantum resonances by approximate quantum numbers becomes questionable already above the third Landau threshold.

We conclude, that the gross structure of the spectrum is reproduced by our new QEBK – quantization. To resolve the fine structure in the spectrum and the additional resonances, we have to include the chaotic part of the dynamics. This can be done in a systematic way by performing the cycle expansion (37) including higher order curvature terms. We will present results in the next section.

The QEBK– approach is in particular useful to derive analytic expressions like equation (56) close to the Landau thresholds. In order to demonstrate the accuracy of the QEBK formulas in the large $M$ limit, we insert the real part of the quantum eigenvalues from Table 2 – 4 on the right hand side of (56) and replace $S_0^1$ on the left hand side by a variable $S_{N,M}$ to fulfill the equality sign. The semiclassical theory then predicts $\lim_{M\to\infty} S_{N,M}$ equals a constant, which is independent of $N$ and the limiting value is given by the classical coefficient $S_0^1 = -0.6483$. Results for different

|   |   | Quantum Mechanics | | QEBK | | Cycle Expansion | | |
|---|---|---|---|---|---|---|---|---|
| $N$ | $M$ | Re $s$ | Im $s$ | Re $s$ | $\delta$ | Re $s$ | Im $s$ | $\delta$ |
| 1 | 0 | 1.35035 | -0.08586 | 1.44120 | 17.04 | 1.33443 | -0.08332 | 2.90 |
| 1 | 1 | 1.79920 | -0.08930 | 1.95646 | 43.55 | 1.78245 | -0.08668 | 4.24 |
| 1 | 2 | 2.03520 | -0.01729 | 2.22902 | 74.38 | 2.02967 | -0.01435 | 1.83 |
| 2 | 0 | 2.25324 | -0.00009 | 2.32694 | 13.65 | 2.25150 | -0.00084 | 0.32 |
| 1 | 3 | 2.41661 | -0.00480 | 2.40817 | 5.43 | 2.41209 | -0.00241 | 2.93 |
|   |   | 2.44319 | -0.06541 |   |   | 2.45229 | -0.05027 | 6.38 |
| 1 | 4 | 2.53989 | -0.00063 | 2.53346 | 5.76 | 2.53937 | 0.00010 | 0.47 |
| 1 | 5 | 2.62677 | -0.4 $10^{-6}$ | 2.62435 | 2.92 | 2.62558 | 0.00076 | 1.44 |
| 1 | 6 | 2.69279 | -0.00010 | 2.69213 | 1.06 | 2.69172 | 0.00025 | 1.71 |
| 1 | 7 | 2.74367 | -0.00020 | 2.74381 | 0.27 | 2.74287 | -0.00005 | 1.68 |
| 1 | 8 | 2.78347 | -0.00023 | 2.78396 | 1.30 | 2.78289 | -0.00015 | 1.57 |
| 1 | 9 | 2.81504 | -0.00022 | 2.81568 | 2.14 | 2.81461 | -0.00016 | 1.45 |
| 1 | 10 | 2.84043 | -0.00019 | 2.84111 | 2.88 | 2.84010 | -0.00013 | 1.37 |
| 1 | 11 | 2.86108 | -0.00015 | 2.86177 | 3.56 | 2.86082 | -0.00009 | 1.33 |
| 1 | 12 | 2.87808 | -0.00011 | 2.87875 | 4.23 | 2.87786 | -0.00005 | 1.34 |
| 1 | 13 | 2.89221 | -0.00008 | 2.89287 | 4.93 | 2.89202 | -0.00001 | 1.39 |
| 1 | 14 | 2.90406 | -0.00005 | 2.90470 | 5.75 | 2.90389 | 0.00003 | 1.45 |
| 1 | 15 | 2.91405 | -0.00006 | 2.91472 | 7.03 | 2.91393 | 0.00005 | 1.24 |
| 2 | 1 | 2.95568 | -0.02187 | 2.96408 | 2.09 |   |   |   |
| 1 | $\infty$ | 3.0 | 0.0 | 3.0 |   | 3.0 | 0.0 |   |

Table 3: The same as Table 2 for $1 < Re(s) \leq 3$.

|   |   | Quantum Mechanics | | QEBK | | Cycle Expansion | | |
| --- | --- | --- | --- | --- | --- | --- | --- | --- |
| $N$ | $M$ | Re $s$ | Im $s$ | Re $s$ | $\delta$ | Re $s$ | Im $s$ | $\delta$ |
| 3 | 0 | 3.17878 | -0.05284 | 3.21081 | 5.92 | 3.17005 | -0.05350 | 1.60 |
| 2 | 2 | 3.32848 | -0.02099 | 3.34358 | 4.81 | 3.32850 | -0.02063 | 0.01 |
| 2 | 3 | 3.57591 | -0.02234 | 3.62001 | 17.51 | 3.57228 | -0.02357 | 1.41 |
|   |   | 3.73250 | -0.01734 |   |   | 3.73010 | -0.02003 | 1.09 |
| 2 | 4 | 3.84891 | -0.02636 | 3.83275 | 8.26 | 3.85544 | -0.02964 | 3.38 |
| 3 | 1 | 3.97355 | -0.04542 | 3.92706 | 10.76 |   |   |   |
| 2 | 5 | 4.00158 | -0.00995 | 4.00118 | 0.25 | 3.98719 | -0.00183 | 8.92 |
| 4 | 0 | 4.09806 | -0.03383 | 4.09410 | 0.73 | 4.06406 | -0.03421 | 6.25 |
| 2 | 6 | 4.18873 | -0.01770 | 4.13706 | 41.36 | 4.17664 | -0.02466 | 10.01 |
| 2 | 7 | 4.27474 | -0.00251 | 4.24824 | 25.29 | 4.27393 | -0.00900 | 0.79 |
| 2 | 8 | 4.34968 | -0.00062 | 4.34024 | 10.68 | 4.35543 | -0.00342 | 6.61 |
| 2 | 9 | 4.41610 | -0.01134 | 4.41713 | 1.38 | 4.42244 | 0.00335 | 8.52 |
| 3 | 2 | 4.43167 | -0.00953 | 4.38177 | 13.95 |   |   |   |
| 2 | 10 | 4.48467 | -0.00385 | 4.48195 | 4.32 | 4.47754 | -0.00569 | 11.26 |
| 2 | 11 | 4.53808 | -0.00585 | 4.53701 | 1.98 | 4.53703 | -0.01022 | 1.95 |
| 2 | 12 | 4.58573 | -0.00697 | 4.58412 | 3.53 | 4.58751 | -0.00905 | 3.89 |
| 2 | 13 | 4.62807 | -0.00650 | 4.62468 | 8.64 | 4.62954 | -0.00631 | 3.78 |
|   |   | 4.63908 | -0.06772 |   |   |   |   |   |
| 2 | 14 | 4.66388 | -0.00396 | 4.65982 | 11.96 | 4.66431 | -0.00336 | 1.28 |
| 2 | 15 | 4.69294 | -0.00201 | 4.69042 | 8.48 | 4.69296 | -0.00161 | 0.07 |
| 2 | 16 | 4.71804 | -0.00210 | 4.71722 | 3.12 | 4.71809 | -0.00179 | 0.22 |
| 2 | 17 | 4.74127 | -0.00289 | 4.74081 | 2.01 | 4.74126 | -0.00235 | 0.06 |
| 3 | 3 | 4.76087 | -0.05928 | 4.72799 | 10.92 |   |   |   |
| 2 | 18 | 4.76228 | -0.00318 | 4.76164 | 3.13 | 4.76219 | -0.00254 | 0.43 |
| 4 | 1 | 4.86557 | -0.01118 | 4.86649 | 0.20 |   |   |   |
| 5 | 0 | 4.93094 | -0.15662 | 4.97714 | 8.45 |   |   |   |
| 2 | $\infty$ | 5.0 | 0.0 | 5.0 |   | 5.0 | 0.0 |   |

Table 4: The same as Table 2 for $3 < Re(s) \leq 5$.

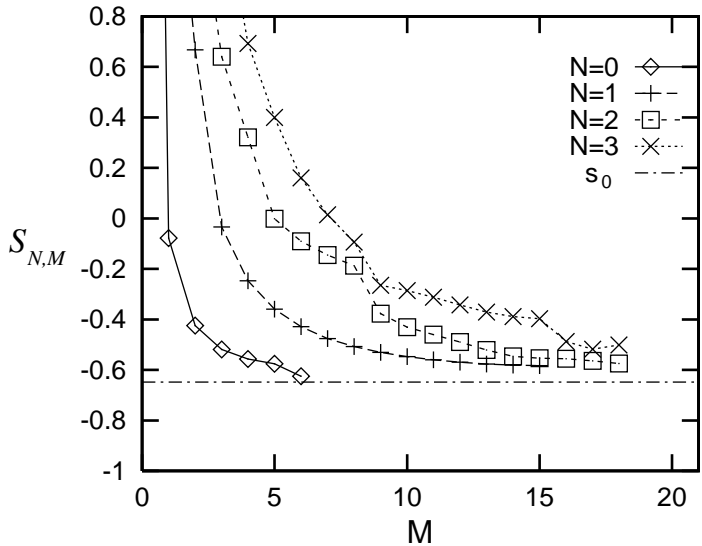

Figure 5: The quantity $S_{N,M}$ for the different $N$ – Landau channels versus the $M$ – Rydberg quantum number for the $m^\pi = 0^+$ and $\epsilon = 0.5$. The dashed dotted line corresponds to the semiclassical limit $S_0^1 = -0.6483\ldots$.

Landau channels and $\epsilon = 0.5$ are presented in Fig. 5 and support our conjecture in detail.

We conclude that the spectrum of the DKP is purely integrable deep in the Landau channel with two important modifications; the Maslov phase for the motion along the magnetic field axis is non integer and the spectrum depends on an additional parameter $S_0^1(\epsilon)$ which can be determined by the classical dynamics.

The QEBK–approximation allows for a quantization of both the real and imaginary part of $s$. The quantization condition for $Im(s)$ can be obtained by inserting the quantized winding numbers $x_{N,M}$ and energies $u_{N,M}$ in (48) and (49) which yields the asymptotic expressions for $v = Im(s)$,

$$v_{N,M} = -\frac{1}{4\pi\epsilon^4}\left(l_0^{(1)} + \log(\frac{N+1/2}{3\epsilon})\right)\left(\frac{N+1/2}{M+1/8}\right)^3, \quad M \gg 1. \tag{57}$$

A decrease in the width of the resonances proportional to $M^{-3}$ is indeed typical for atomic systems [55].

The antisymmetric spectrum $\pi_z = -1$ is obtained by a sign change in (49) or equivalent by replacing the phase factor $1/8$ by $5/8$ throughout this section. The comparison between the QEBK quantization and exact quantum calculations yields results of the same quality as for the $\pi_z = +1$ spectra.

## 3.3 Cycle expansion results

So far, we can describe the gross structure of the spectrum in terms of a quasi integrable approximation, which can be derived from the fundamental term $c_1$ in the cycle expansion (37). The finer details in the spectrum, as e.g. level repulsion due to the overlap and interference of different Rydberg series is a manifestation of the non–integrability of the system and can not be resolved in the QEBK–approach. The non–integrability enters the semiclassical formalism through the chaotic dynamics in the near core region and its coupling to the regular channel. The full information about this part of the classical dynamics is contained in the higher curvature terms $c_n$ in the cycle expansion (37). An analytic continuation of the multiple infinite sums below the real $s$ axis leads to multi dimensional infinite integrals as discussed in detail in section 3.1. The one dimensional integrals (43) in the fundamental term $c_1$ can be calculated "exactly" by standard numerical techniques. A full treatment of the higher curvature terms demands a determination of periodic orbit dependent coefficient matrices in the expansion of the actions and stability exponents like in Eqs. (44), and an evaluation of higher dimensional integrals. At present, this exceeds the limits of our computational abilities.

| length in the $\infty$–alphabet | upper index | lower index (up to permutation) | |
|---|---|---|---|
| 2 | 11, 12, 22 | $(n, m);$ | $n = 1, \ldots \infty$ $m \leq 3$ |
| 3 | 111, 112, 122, 222 | $(n, m, k);$ | $n = 1, \ldots \infty$ $m + k \leq 3$ |
| 4 | 1111, 1112, 1122, 1212, 1222, 2222 | $(n, m, k, l);$ | $n = 1, \ldots \infty$ $m + k + l = 3$ |

Table 5: Cutoff of the tensor summations in the cycle expansion (37).

Instead, we approximate the higher order curvature contributions in (37) by carrying out the analytic continuation only in one of the summation indices and sum over the first few numbers in the other indices directly. (Summing over the lower indices $m$, $k$ or $l$ in Table 5 up to infinity leads of course to the same divergences for $Im(s) < 0$ as described for the single sums in chapter 3.1. This is, however, no problem as long as $m$, $k$, $l$ is small.) The treatment is equivalent to a cycle expansion in terms of periodic orbit families of the form $G\,0^{n-1}$, where the head string $G$ is here a substring of length $n_G$ in our old ternary alphabet starting and ending with a '1' or '2'. The results presented below belong to a cycle expansion including all periodic orbit families up to a maximal head length $n_G = 4$. The corresponding symbols in the infinite alphabet are listed in Table 5, the leading order family dependent coefficients in the expansion of the actions and stabilities are included in Table 1. Our expansion of the zeta function thus contains 54 periodic orbit families and it includes contributions up to the $4^{\text{th}}$ curvature term in the cycle expansion (37).

Fig. 6 shows the absolute value of the cycle expanded zeta function for $Im(s) = 0$ together with the real parts of the quantum resonances marked as ticks on the $Re(s)$ – axis. The minima of the zeta function coincide very well with the quantum spectrum and the Rydberg series structure is reproduced leading to an essential singularity of the zeta function at each Landau threshold $Re(s) = (N + 1/2)/\epsilon$. The perturber states appear as strong modulations in the Rydberg – oscillation, (indicated by the arrows in Fig. 6). To obtain both the real and imaginary part of the quantum eigenvalues, we calculated the zeros of the zeta function in the complex $s$ – plane. Results are shown in Fig. 7 in comparison with the complex eigenvalues obtained from our quantum calculations. The numerical values are listed in Tables 2–4. Our semiclassical zeta function indeed contains the information about both the position and width of the resonances. The strong oscillations in the imaginary parts of the eigenvalues are well reproduced and the error is of the same order of magnitude as for the real part. We also resolve the first states missing in the QEBK approach at $Re(s) = 2.443$ and $3.733$. Also the ground states of the $N = 2, 3, 4$ series perturbing the lower Rydberg series are now obtained including level repulsion effects! The error $\delta$ in units of the mean level spacing is uniform in the neighborhood of these perturbing states. In addition, the overall deviation of the semiclassical eigenvalues from the quantum results decreases compared to the QEBK–approach. A remarkable exception is the series of bound states $N = 0$. The real part of the zeros of the zeta function reproduces the quantum data again well, but we obtain (in contrast to all the other Landau series) here a positive imaginary part! A correction of the systematic deviation of these zeros from the real $s$ – axis are still an open question. The occurrence of bound states itself is a pure wave effect, and is associated with the largest wave length possible in the system. We thus expect higher $\hbar$–terms to play a crucial role in the semiclassical expansion of the low lying part of the spectrum, here the $N = 0$ bound states. Similar effects have been observed in scattering billiards [56, 57], where the next to leading $\hbar$ corrections have been calculated explicitly.

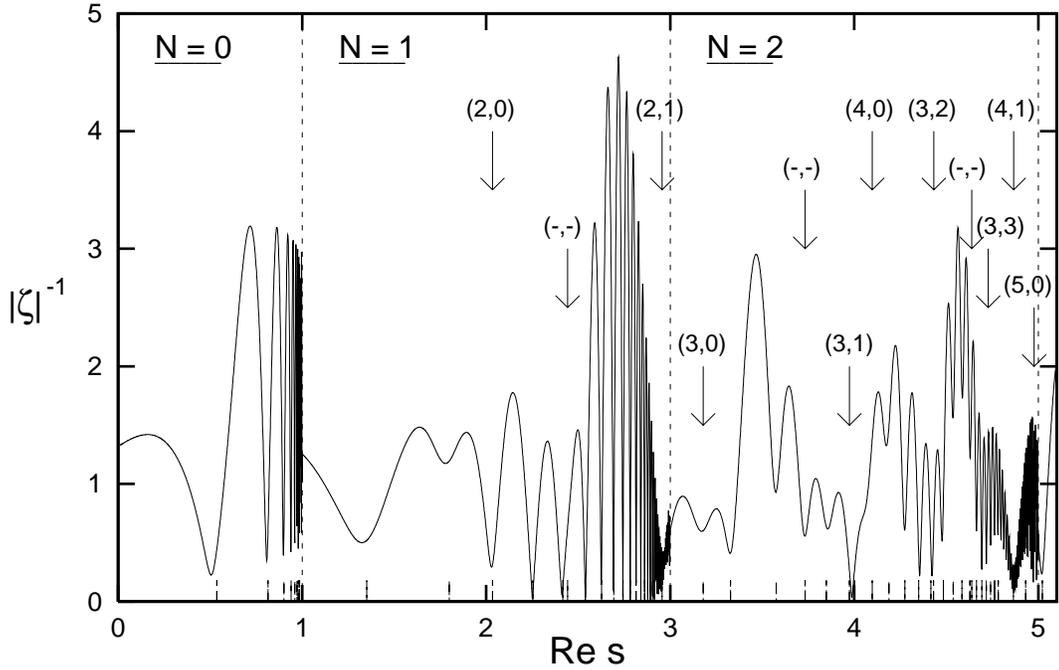

Figure 6: The absolute value of the cycle expanded zeta function for the $m^\pi = 0^+$ subspace and $\epsilon = 0.5$; the ticks on the $s$ axis denotes the position of the quantum resonances and the arrows indicate perturber states due to the overlap of different Rydberg series. (The quantum numbers (N,M) are obtained from Table 2–4.)

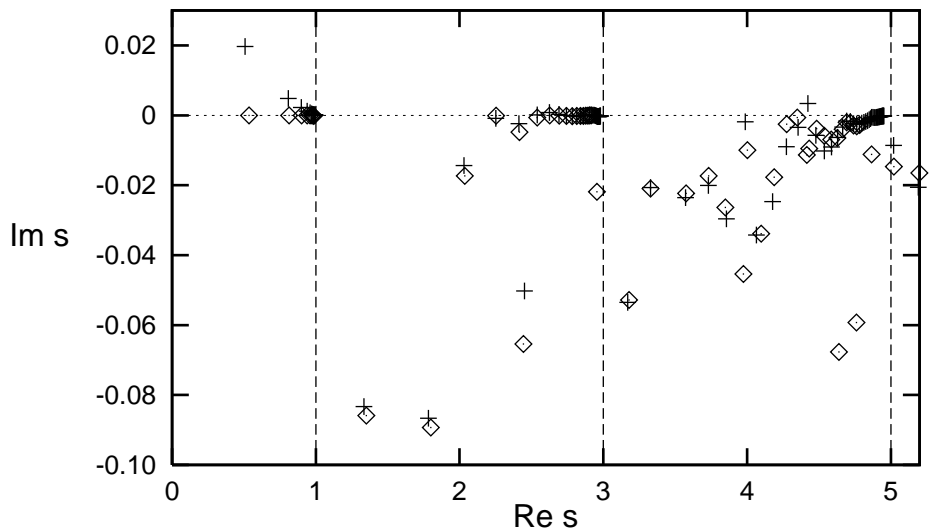

Figure 7: The zeros of the cycle expanded zeta function (+) compared with the quantum results (◇) for $m^\pi = 0^+$ and $\epsilon = 0.5$ plotted in the complex $s$ plane.

Our cycle expansion of the zeta function can so far not resolve all quantum states, zeros are missing near strong perturbations of the Rydberg series, see Tables 3, 4. (Note, that these perturber can still be observed as modulations in the zeta function, Fig. 6). The reason for not reproducing all the quantum eigenstates is well understood and not a principle problem of the theory. A semiclassical quantization of a generic system with more than one degree of freedom demands periodic orbits of increasing length or equivalent higher curvature terms to resolve the increasing level density in the spectrum or in our example the increasing number of perturbing resonances. The number of orbits itself grows exponentially with the period, which usually sets numerical limits to the maximal curvature contributions attainable [58].

In addition, we approximate here the curvature terms itself by the shortest periodic orbit families only. We miss thereby information about the dynamics in the classical Landau channels by periodic orbits visiting the far regular regions two or more times. These orbits build up perturbations of the Rydberg series near the Landau thresholds due to the overlap of different $N$ – series. As a consequence, we can not resolve perturber states with large imaginary part in the Rydberg series for large $M$ quantum numbers. Both effects can be read off from Fig. 8, where we plotted the modulus of the curvature contributions for $Im(s) = 0$. The curvature contributions drop off exponentially for fixed $s$, and we obtain a decrease over 3 decades for $Re(s)$ close to zero! The total fall off becomes smaller inside each $N$ – interval, (but is still exponential) and decreases also globally with increasing $N$. The non uniform behavior of the various curvature contributions in each Rydberg series is due to the finite approximation of the curvature terms itself, the overall increase is a manifestation of the cutoff in the curvature expansion.

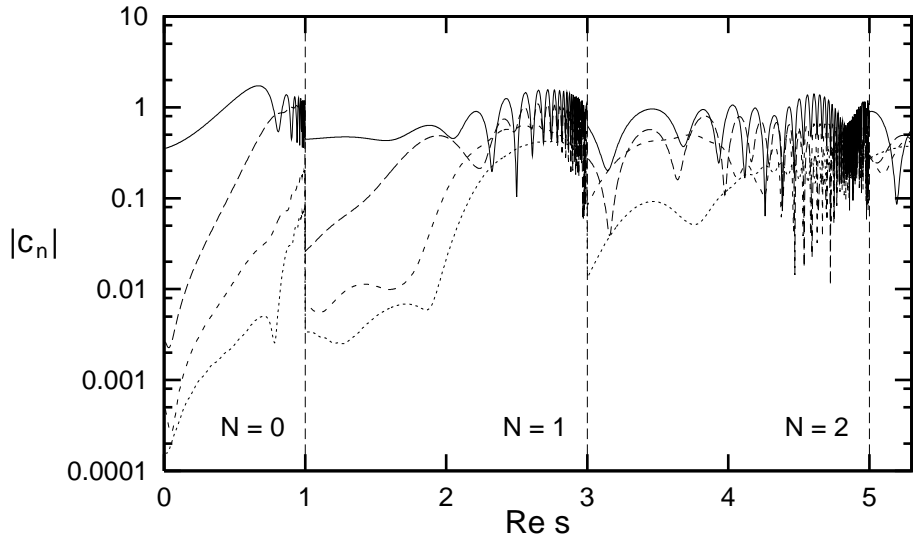

Figure 8: The absolute value of the first 4 curvature contributions $c_n$ of the zeta function for real $s$ – values.

We conclude, that the cycle expanded zeta function with a double infinite symbolic dynamics includes all the important features of the whole spectrum. We resolve both the Rydberg – like structure at the Landau levels and the perturbations due to the overlap of the different $N$ – series. The deviations from the quantum results and the missing levels can be understood in the context of the approximations made in the cycle expansion. A quantization of the odd parity spectrum $m^\pi = 0^-$ is obtained by changing the Maslov index according to (34) and yields results of the same quality as presented above. The semiclassical quantization presented here thus exceeds standard quantum mechanical methods in principle, which have problems to obtain the high lying Rydberg resonances in the continuum, see [59] for an alternative approach.

Our results also shed a new light on recent speculations about the origin of the very narrow resonances or quasi–bound states, which have been observed in Hydrogen in a magnetic field [60, 18], (see e.g. the eigenvalue at $Re(s) = 2.627$ in Table 3.) Our semiclassical treatment suggests, that these extremely small imaginary parts are due to pure interference effects. They can not be traced back to special features in the classical dynamics, as e.g. to stable classical motion, but can be reproduced as a collective effect in a quantization using unstable periodic orbits only. This is in contrast to the Helium atom, where series of quasi–bound resonances could be assigned to a large stable island in phase space [54]. As a consequence, these small resonances in the DKP cannot be expected to be correlated with the classical relevant lines $\epsilon = const$. Detailed numerical studies indeed found a more or less 'random' distribution of narrow resonances over the whole $E - \gamma$ parameter plane [61].

# 4   Conclusions

We have presented here the first semiclassical calculations of the resonance spectrum in the Diamagnetic Kepler problem. A proper treatment of the system can not be performed without overcoming fundamental problems in the semiclassical approach originating from the intermittency in the classical dynamics. A regularisation of the divergences in the semiclassical zeta function caused by the regular part of the classical motion is obtained by performing the cycle expansion of the zeta function in terms of a double infinite alphabet. The fundamental term and the curvature contributions are written as infinite sums over families of periodic orbits. The orbits in these families approach the marginal stable fixed point in the separable limit of the potential. An analytically continued integral representation of the various curvature contributions is given. The universal asymptotic behavior of the actions and stability exponents in the different periodic orbit families could be derived analytically, subleading contributions are obtained from periodic orbit data directly.

An optimized version of the complex rotation technique is presented, which allow for a calculation of the resonance spectrum for relatively high lying Rydberg states at the various Landau thresholds.

The good agreement between "exact" and semiclassical results both in the position and in the width of the resonances demonstrates that the semiclassical zeta function as the leading term in an $\hbar$ – expansion indeed carries most information about the spectrum. An approximation of the zeta function itself uncovers an EBK–like quantization of the marginal stable fixed point at infinity. This quantization scheme may serve as a useful tool to obtain an overall estimate of the spectrum and yields surprisingly accurate results especially for the ground state series and for the Rydberg spectrum near the Landau thresholds. Our results suggest, that the extremely narrow resonances in the DKP are a pure quantum interference effect, (which is reproduced qualitatively by our semiclassical method using unstable periodic orbits), and are not associated with stable classical motion.

Our results so far may serve as a basis for further projects such as the study of $\hbar$–corrections. Of fundamental interest is a deeper understanding of the semiclassical theory, when crossing the critical scaled energy $\epsilon_c$ from above. This is no problem for pure quantum techniques, but complicates semiclassical methods as presented here significantly due to bifurcations of periodic orbits and pruning in the symbolic description of the dynamics.

**Acknowledgement**
We thank Predrag Cvitanović, John S. Briggs, Jan–Michael Rost, and Gernot Alber for constructive discussions. G.T. would like to thank the Deutsche Forschungsgemeinschaft for financial support, K.T. is grateful for financial support from the Alexander von Humboldt – Stiftung and the Norwegian Research Council (NFR), and J.M. appreciate support from the Deutsche Forschungsgemeinschaft (SFB 237).

# A  Symmetry reduced symbolic dynamics

A three letter symbolic alphabet for the 4–disk problem and the diamagnetic Kepler problem is obtained by defining the new symbols as the relative increment of two consecutive bounces $s_t s_{t+1}$ with two disks. Our choice of a ternary symbolic dynamics is a symbolic description which is closely related to the $C_{4v}$ – symmetry in the 4–disk billiard and in the DKP – Hamiltonian (4) in $\mu, \nu$ – coordinates. The new symbols correspond to the symmetry operations forming the $C_{4v}$ group. Orbits, which can be mapped onto each other by symmetry operations, will have the same code, here. The symbolic dynamics thus describes the motion in the desymmetrized fundamental domain, (which is e.g. the area $\mu \geq \nu \geq 0$ in Fig. 2), with hard reflection at the boundaries. The symbolic description was first introduced by Cvitanović and Eckhardt [33] in the 4–disk problem.

We denote this symbolic dynamics $G = \cdots g_{-2} g_{-1} g_0 g_1 g_2 \cdots$ with $g_i \in \{0, 1, 2\}$. The disk enumerating symbols $S$ are obtained from the symbols $G$ by the following algorithm: Choose one disk as the starting disk $s_1$ (e.g. $s_1 = 1$) and the next disk is given as $s_2 = s_1 + 1 \mod 4$ if $g_1 = 0$ or $g_1 = 1$ and it is $s_2 = s_1 + 2 \mod 4$ if $g_1 = 2$. If the symbol $g_1$ was a 1 or a 2 we continue adding 1 or 2 to obtain $s_3$ depending on the value of $g_2$. If $g_1$ was a 0 we change the sign of the increment for the next symbol and choose the decrement $-1$ or $-2$ to get $s_3$. Each symbol $g_i = 0$ change the sign of the increment for the following symbols while $g_i = 1$ preserves the sign.

Using this rule we obtain for the periodic orbit $G = \overline{2}$ the strings $S = \overline{13}$ or $S = \overline{24}$; from $G = \overline{1}$, $S = \overline{1234}$ or $S = \overline{1432}$; from $G = \overline{20}$, $S = \overline{1324}$, $S = \overline{1342}$, $S = \overline{1423}$, or $S = \overline{1243}$; and from $G = \overline{10}$, $S = \overline{1232}$, $S = \overline{2343}$, $S = \overline{3414}$, or $S = \overline{4121}$. These are the four orbits drawn in Fig. 2.

In the alphabet $G$ we have the same symbolic description for all periodic orbits which can be mapped into each other with a spatial rotation or a reflection. Cycles which are related to each other by time reversal symmetry alone will be represented by two different symbol strings.

The symmetry of a periodic orbit is obtained by identifying each symbol with a discrete group operator of the $C_{4v}$ group and multiply the elements of the symbol string describing the cycle together. We identify symbol 0 with the reflection $\sigma$ on the $\mu$ or $\nu$ axis, symbol 1 with a $\pi/2$ – rotation or a $C$ symmetry, and symbol 2 with $C_2$, a $\pi$ – rotation of the configuration space. The total symmetry of the periodic orbit becomes important, when looking at the periodic orbit zeta function in its symmetry factorized representation.

The zeta-function for the 4-disk system factories according to the irreducible representations of the group $C_{4v}$ in $\zeta = \zeta_{A_1} \zeta_{A_2} \zeta_{B_1} \zeta_{B_2} \zeta_E$. The factorization of the single orbit contributions can be read off from Ref. [33]. The DKP in $\rho, z$ coordinates, however, has only $C_2$ symmetry due to the invariance of the Hamiltonian (2) under the transformation $z \to -z$. The symmetric and antisymmetric representation of the group $C_2$ correspond here to the $A_1$ and $B_1$ representation of the $C_{4v}$ symmetry of the regularized Hamiltonian (4). The DKP – zeta function can thus be written as product $\zeta = \zeta_{A_1} \zeta_{B_1}$. The zeros of $\zeta_{A_1}$ yields the symmetric eigenstates, while $\zeta_{B_1}$ gives the antisymmetric part of the spectrum with respect to parity transformation. The periodic orbits contributing to the single factors in the zeta function are those in the fundamental domain.

Symmetric states belong to a quantization of the DKP in the fundamental domain with Neumann boundaries along the line $z = 0$, antisymmetric states obey Dirichlet boundary conditions there. In the zeta function $\zeta_{B_1}$, a periodic orbit picks up a phase $\pi/2$ every time it crosses the axis $z = 0$. The number of crossings is equal to the number of 1's in the code. A periodic orbit with an odd number of symbols '1' thus changes sign in the zeta function $\zeta_{B_1}$. In addition, the cycle 2, which runs along the symmetry line $z = 0$, does not contribute to an antisymmetric eigenstate. These states have a node on the symmetry axis, the orbit 2 is excluded in the $B_1$ – zeta function.

A time shift corresponds to a cyclic permutation in the symbol code $g_i$. Note however, that a time reversed symbol string corresponds not to the time reversed physical orbit.

# B  Well ordered symbolic dynamics

A second useful ternary symbolic dynamics is the one which has a natural ordering identical to the ordering of the folds in the stable and unstable manifolds. This enables us to make a bisection search in the Poincaré map to find periodic orbits using the method introduced in [34]. Well ordered symbolic dynamics $W = \cdots w_{-2} w_{-1} w_0 w_1 w_2 \cdots$ for the 4 disk system is defined in Ref. [37]. The symbol $w_t \in \{0, 1, 2\}$ is obtained by the increment $s_{t+1} - s_t - 1$ for odd time $t$ and the decrement $s_t - s_{t+1} - 1$ for even time $t$. Reading the string with $t > 0$ as the number $\gamma = \sum_{t=1}^{\infty} w_t / 2^t$ gives the relative position of the fold in the stable manifold structure and equivalent for negative time giving the position in the unstable manifold structure.

We use this symbolic position for points on a Poincaré plane to search for the periodic orbits. We have been able to find orbits up to symbol length 1000 with this method and we have for each family of orbits calculated all orbits up to length 40 and some orbits up to length 500.

Note, that for the well ordered symbols $w_t$ a time shift is not a simple shift operation in the symbol string because the symbols depends on whether the time is even or odd.

# C  The trace of the stability matrix in the classical Landau channel

The stability matrix $\mathbf{M}$, Eq. (21), of a periodic orbit going far out in the classical Landau channel $|z| \gg \rho$ in (2) can be split into two parts; by fixing an arbitrary, but large $|z_0|$, we write

$$\mathbf{M} = \mathbf{M}_0 \, \mathbf{M}_{sep}, \tag{58}$$

where the matrix $\mathbf{M}_{sep}$ describes the linearized motion in the neighborhood of a periodic trajectory in the regular channel from a starting point $(\rho, z_0, p_\rho, p_{z_0} > 0)$ to the end point $(\rho', z_0, p'_\rho, p'_{z_0} \approx -p_{z_0} < 0)$. The chaotic part of the dynamics in the near core region is contained in $\mathbf{M}_0$. The matrix elements of $\mathbf{M}_0$ approach constants, when increasing the number $n$ of oscillations in the regular channel for periodic orbits in the same families. The dominant contribution to the stability matrix $\mathbf{M}_{sep}$ in the regular region can be described by the linearized dynamics of the separable Hamiltonian (9). Transforming our coordinate system from phase space coordinates $(\rho, z)$ to action angle variables $(J, \varphi)$, we can write $\mathbf{M}_{sep}$ as

$$\mathbf{M}_{sep} = \mathbf{T}'^{-1} \tilde{\mathbf{M}}_{sep} \mathbf{T}, \tag{59}$$

where $\mathbf{T}, \mathbf{T}'$ denotes the transformation matrix

$$\mathbf{T} = \frac{\partial(\varphi_h, \varphi_c, J_h, J_c)}{\partial(\rho, z, p_\rho, p_z)}, \tag{60}$$

at the start and ending points. The matrix $\tilde{\mathbf{M}}_{sep}$ is the Jacobi matrix of the flow with respect to action–angle variables and has now a particular simple form,

$$\tilde{\mathbf{M}}_{sep}(t) = \begin{pmatrix} 1 & 0 & 0 & 0 \\ 0 & 1 & 0 & a_c t_c \\ 0 & 0 & 1 & 0 \\ 0 & 0 & 0 & 1 \end{pmatrix} \quad \text{with} \quad a_c = \frac{\partial^2 H_{sep}}{\partial J_c^2} = -3 J_c^{-4}, \tag{61}$$

and $t_c$ is the time spend in the regular channel. Note, that $J_c$ and $t_c \sim J_c^3$ depends through (13) on the energy deposed in the Coulomb motion. The transformation $\mathbf{T}$ can be given as

$$\mathbf{T} = \begin{pmatrix} \partial \varphi_h / \partial \rho & 0 & \partial \varphi_h / \partial p_\rho & 0 \\ 0 & \partial \varphi_c / \partial z & 0 & \partial \varphi_c / \partial p_z \\ \omega \rho & 0 & p_\rho / \omega & 0 \\ 0 & J_c^3 / z^2 & 0 & J_c^3 p_z \end{pmatrix}. \tag{62}$$

Using the condition $\det \mathbf{T} = 1$, we can write down a similar expression for the inverse of $\mathbf{T}$. The derivatives of the angle variables approach finite values for periodic orbits in the same family and $n \to \infty$. The explicit expressions for the derivatives are not needed here.

Inserting (61) and (62) in (58), we can approximate the trace of $\mathbf{M}$ as follows,

$$\text{Tr} \mathbf{M} \approx k_1 J_c^2 t_c + k_2 J_c^3 + k_3. \tag{63}$$

All the expressions, which approach constants in the limit $t_c \to \infty$ or $E_c \to 0$, are collected in $k_1$, $k_2$, $k_3$. The time variable $t_c$ is proportional to the winding number $\alpha$ and $n$, and using Eqs. (16), (19), we obtain $J_c \sim n^{1/3}$. The trace can thus be written as

$$\text{Tr} \mathbf{M} = k_1 n^{5/3} + k_2 n + \mathcal{O}(n^{2/3}) \tag{64}$$

The coefficients $k_i$ depend on the history of the periodic orbit before entering the regular region and on the scaled energy $\epsilon$. The trace of $M$ is independent of $\gamma$ after one period of the periodic orbit.